\documentclass[aps,pra,twocolumn,superscriptaddress,nofootinbib,showpacs]{revtex4-1}

\usepackage{graphicx}
\usepackage{color,ulem}
\usepackage{amsmath}
\usepackage{amssymb}
\usepackage{amsfonts}

\graphicspath{{Plots/}}
\begin{document}

\title{Tunneling decay of two interacting bosons in an asymmetric double-well potential:\\
	A spectral approach
 }

\author{Stefan Hunn}
\affiliation{Physikalisches Institut, Albert-Ludwigs-Universit\"at Freiburg, Hermann-Herder-Str. 3, 79104 Freiburg, Germany}
\author{Klaus Zimmermann}
\affiliation{Physikalisches Institut, Albert-Ludwigs-Universit\"at Freiburg, Hermann-Herder-Str. 3, 79104 Freiburg, Germany}
\author{Moritz Hiller}
\affiliation{Physikalisches Institut, Albert-Ludwigs-Universit\"at Freiburg, Hermann-Herder-Str. 3, 79104 Freiburg, Germany}
\affiliation{Institute for Theoretical Physics, Vienna University of Technology, Wiedner Hauptstra{\ss}e 8-10/136, 1040 Vienna, Austria}
\author{Andreas Buchleitner}
\affiliation{Physikalisches Institut, Albert-Ludwigs-Universit\"at Freiburg, Hermann-Herder-Str. 3, 79104 Freiburg, Germany}

\date{\today}

\begin{abstract}
We study the full-fledged microscopic dynamics of two interacting, ultracold bosons in a one-dimensional double-well potential, through the numerically exact diagonalization of the many-body Hamiltonian.
With the particles initially prepared in the left well, we increase the width of the right well in subsequent trap realizations and witness how the tunneling oscillations evolve into particle loss.
In this closed system, we analyze the {\it spectral} signatures of single- and two-particle tunneling for the entire range of repulsive interactions. 
We conclude that for comparable widths of the two wells, pair-wise tunneling of the bosons may be realized for specific system parameters.
In contrast, the decay process (corresponding to a broad right well) is dominated by uncorrelated single-particle decay.
\ \\
\end{abstract}
 
\pacs{%
03.75.Lm, % BEC: -tunneling
03.65.Xp,  % quantum mechanics of tunneling
05.30.Jp  % Boson systems
}

\maketitle

%%%%%%%%%%%%%%%%%%%%%%%%%%%%%%%%%%%%%%%%%%%%%%%%%%%%%%%%%%%%%%%%%%%%%%%%%
%%%%%%%%%%%%%%%%%%%%%%%%%%%%%%%%%%%%%%%%%%%%%%%%%%%%%%%%%%%%%%%%%%%%%%%%%

\section{Introduction}

One of the hallmarks of quantum mechanics is tunneling through energy barriers that classically cannot be overcome.
It is at the heart of fundamental dynamical phenomena like Josephson oscillations \cite{Josephson:1962}, the spatial flipping of the nitrogen atom in the ammonium molecule  \cite{Hund27} or quantum lattice dynamics \cite{SEG94}.
Tunneling is also one of the major mechanisms that provoke the loss of particles from quantum systems \cite{Gamow28}. 
The latter occurs whenever a quantum system is coupled via a potential barrier to a continuum, i.e., to a set of asymptotically free states.
Such particle loss is omnipresent in nature, e.g., in radioactive decay or in the autoionization of excited atomic states.

While the tunneling and loss dynamics are thoroughly understood for single particles \cite{Razavy:2003}, much less is known about the generic case of {\it interacting} many-body systems.
Furthermore, in the vast majority of physical realizations, the inter-particle interactions are immanent and hard to manipulate if it can be done at all.
An archetypical example is the inter-electronic Coulomb interaction which plays a crucial role in the non-sequential double ionization of helium \cite{WSDASK94}, or in the sequential double ionization of argon \cite{Pfeif11}.

In this respect, the advent of ultracold atoms in optical potentials prepared the stage for a new generation of experiments in which all relevant control parameters can be controlled essentially at will, ranging from the potential landscape \cite{MSHCR05,HRMB09,EWARWD10}, over the initial state (and number) of atoms \cite{SZLOWS11}, to the inter-particle interaction, which can be tuned over several orders of magnitude from attractive to repulsive \cite{HGM09,CGJT10}. 
Consequently, several authors have studied the decay of particles from a Bose-Einstein condensate (for recent experiments, see, \cite{Geri_etal08,SBK10}).
The two major theoretical tools used in this context are the mean-field (Gross-Pitaevskii) description \cite{MCMB04,KW04,WMK05,SW07,DFSF09,NHFKG09,Braz_etal09},
and effective Hamiltonian or master equation approaches \cite{HKO06,GKN08,WTW_PRL08,Diehl_etal10,WTHKGW11}, to name a few of them.
While both methods can yield valuable insight, they face fundamental limitations:
The mean-field treatment (applicable in the limit of large particle numbers)
is based on the assumption that the condensate is in a coherent state and breaks down \cite{VA01,MHUB06,Chuc10,BLSACB12}
 when the inter-particle interaction leads to considerable dephasing \cite{BK03,GHMDRN08}.
The master equation, on the other hand, is typically introduced in an {\it ad hoc} procedure without a rigorous justification and few authors went beyond the standard Markovian treatment \cite{BFKP99,MHS99,LNL07}.

In contrast, the {\it microscopic} decay process of interacting particles only begins to be explored on a fundamental level.
Recent experiments reported the interaction-dependent loss of {\it the first} of a few atoms from a one-dimensional trap \cite{ZSLWRBJ12}, and the corresponding escape rate was successfully described by a quasi-particle wave-function approach \cite{Rontani:2012pd}.
The theoretical works that describe the full-fledged decay dynamics rely on the propagation of the time-dependent Schr\"odinger equation 
either by direct integration \cite{KB11}, by combining a matrix-product state approach with Bose-Hubbard-like chains \cite{GC11},  or via multi-configurational Hartree-Fock methods \cite{LSAMC09,LSSAC12}.

In our present work, we take a complementary approach to tunneling decay and study the numerically exact many-body dynamics of two interacting bosons which are initially prepared in the left site of a double-well potential.
In subsequent realizations of the trapping potential, we gradually increase the extension of the right well which ---in the limit of large widths--- mimics the unconfined configuration space (to which the particles escape), by a dense quasi-continuum of states.
Of course, the finite extension of the broad well defines a maximal observation time before the particles hit the rightmost boundary.
This limitation is, however, more than compensated by two major benefits:
On the one hand, we can directly monitor the {\it transition} from the regime of tunneling oscillations between the two wells to the regime of tunneling decay from the left well, in a single setup.
On the other hand, we gain complete access to all spectral quantities of the system (e.g. its eigenvalues, eigenfunctions, and the density of states) thus providing insight to the inner workings of many-body tunneling.

The key problem is to expose the physical nature of the many-body decay process:
Given, for example, the quantum mechanical correlations naturally present in the initial state of two interacting bosons,
do the particles leave the left well as a pair or do they tunnel independently?
That is, depending on the strength of the interatomic interactions, will we observe predominantly correlated or uncorrelated processes?
We would as well like to know which quasi-continuum states actually support the decay process, and how interaction energy is converted into kinetic energy, during the tunneling decay.
To answer these questions, we shall investigate two-body as well as reduced single-particle quantities.

The narrative of this work is as follows:
In the next section we introduce our model, the double-well potential, and those quantities used throughout the paper to study particle loss.
Sec.~\ref{sec.spp} is devoted to the tunneling decay of a {\it single} particle and sets the frame of reference for the two-body problem.
From a spectral perspective, we analyze the transition from tunneling oscillations to tunneling decay as the width of the right well is increased, and will see that the participation number of the initial state in the eigenstates of the double well is an excellent figure of merit to characterize this transition.
We briefly revise single-particle decay theory, and elaborate on the validity of our quasi-continuum approach.

In Sec.~\ref{sec.tp_box} we introduce the initial state of the dynamical two-boson evolution, and define the principal physical observables.
Our central results are presented starting from Sec.~\ref{sec.tp_asdw_spec} where we analyze the spectral properties of the tunneling process of two interacting bosons as the width of the right well is gradually increased.
Specifically, we identify those particle configurations in the quasi-continuum states which support single- and two-particle tunneling, respectively, and calculate the associated density of states.
This procedure parallels the analysis of Sec.~\ref{sec.spp}, and the differences to the single-particle case are exposed.
In Sec.~\ref{sec.tp_asdw_dyn} the predictions of our spectral analysis are compared to the numerically obtained time evolution.
We develop a complete picture of the decay process by means of two- and single-particle quantities and compare our results to related studies on the time-resolved decay dynamics \cite{LSAMC09,KB11,LSSAC12}.
We conclude with a summary and discussion in Sec.~\ref{sec.concl}.

%=== MODEL ===
\section{Double-well potential as a model for tunneling decay}
\label{sec.model}

\begin{figure}[t]
\centering
\includegraphics[ width=0.9\columnwidth,keepaspectratio]{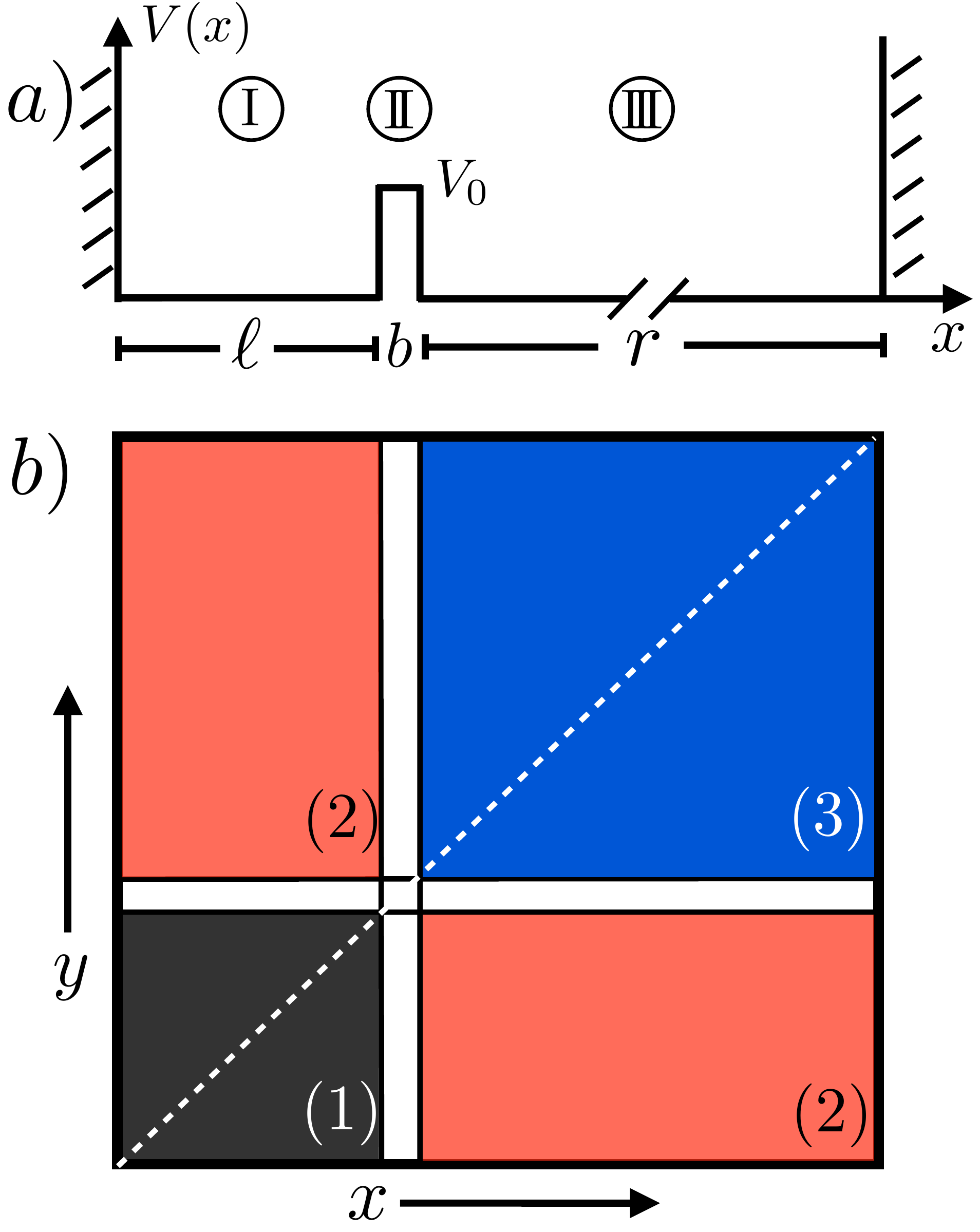}\hfil
\caption{ (Color online) {\it Top:} (a) Schematic sketch of the single-particle potential $V(x)$ as defined in Eq.~(\ref{eq._V_sp}).
 The left well of width ${\ell}$ (interval I) is coupled via a potential barrier of width $b$ and height $V_0$ (interval II) to the right well of width $r$ (interval III).
 {\it Bottom:} (b) Potential landscape of (\ref{eq.H_tp}) in the two-boson $x\!-\!y$ configuration space.
 Two particles confined in the left well correspond to region (1) (black), one particle in either well corresponds to the two
 regions (2) (red or light gray), and both particles in the right well corresponds to region (3) (blue or dark gray).
 The tunneling barrier is depicted in white.
 The dashed white line denotes the diagonal $(x=y)$ at which the contact interaction $U$ between the two particles arises. 
}
\label{fig.setup}
\end{figure}

We now prepare our tools to describe the tunneling dynamics of two ultracold, interacting bosons in a one-dimensional double-well potential.
For $x$ and $y$ the respective coordinates of the (indistinguishable and spinless) bosons, the two-particle Hamiltonian reads
\begin{equation}
H_{tp}(x,y) = H_{sp}(x) + H_{sp}(y) + W(x,y)\, ,
\label{eq.H_tp}
\end{equation}
where $H_{sp}$ is the single-particle Hamiltonian and $W(x,y)$ accounts for the inter-particle interaction (see also Appendix \ref{sec.app}). 
The former is given by
\begin{equation}
H_{sp}(x)=-\frac{{\rm d}^2} {{\rm d}x^2} + V(x) \, ,
\label{eq.H_sp}
\end{equation}
where we have fixed the particle mass ${\tt m}\equiv1/2$ and $\hbar\equiv1$.
The corresponding single-particle potential $V(x)$ is depicted in Fig.~\ref{fig.setup}a):
The left well of width $\ell$ (interval I) is coupled via a potential barrier of width $b$ and height $V_0$ (interval II) to a second well of width $r$ (interval III),
\begin{equation}
V_{}(x)=\begin{cases}
0 & x \in {\rm I}  \mbox{ or }x\in {\rm III} \\
V_0 &x \in {\rm II} \\
\infty &\text{else}
\end{cases} \, .
\label{eq._V_sp}
\end{equation}
The interaction term $W(x,y)$ is assumed to result predominantly from $s$-wave scattering \cite{Legget01}, due to the extremely low temperature of the bosons.
This leads to a $\delta$-like interaction between the particles, i.e., $W(x,y)=U\delta(x-y)$ with $U=2\hbar \omega_{rad} a_s$ \cite{Olshanii98}.
Here, $a_s$ is the $s$-wave scattering length, which can be tuned experimentally over several orders of magnitude, from attractive to repulsive \cite{HGM09,CGJT10}, and $\omega_{rad}$ denotes the frequency of the harmonic {\it radial} trapping potential which acts transversally to the direction of (\ref{eq.H_sp}).
For a cigar-shaped trap (i.e., a small longitudinal trap frequency $\omega_{long}\ll \omega_{rad}$) we can effectively treat the potential as being one dimensional, since ---due to the low total particle energy--- only the transversal ground state is occupied. 
To define the units (of energy etc.), we introduce a reference length scale ${\cal L}$.
With the above convention on $\hbar$ and ${\tt m}$, energy is thus measured in units of $1/{\cal L}^2$, time in units of ${\cal L}^2$, and length in units of ${\cal L}$. The interaction strength $U$ carries the dimension energy times length and is thus measured in units of $1/{\cal L}$.

To investigate tunneling decay (or, synonymously, particle loss) we consider the following dynamical scenario:
Initially, the particles are prepared in the many-body ground state of the {\it isolated left well}, corresponding to an infinitely high tunneling barrier $V_0=\infty$ (or $b=r=0$).
Then, the barrier height is instantaneously reduced to a finite value $V_0=0.1$.
We study the ensuing time evolution for different trap configurations, at fixed widths ${\ell}=51$ and $b=2$, but for ever broader right wells.
The values for $b$ and $V_0$ are chosen such that significant tunneling takes place within comparatively short times.
At the same time, this choice ensures that the energy of the initial state is always considerably smaller than the height $V_0$ of the barrier, such that quantum tunneling is the only way to leave the left well.

In this way, we define a {\it closed} setup, in which the left well serves as the {\it system} while the broad right well possesses a large density of states and can be regarded as the {\it environment} to which the bosons escape.
Throughout this work, environment is to be understood as a quasi-continuum of (two-boson) states, i.e., as the unconfined configuration space.
The rational behind this approach is twofold:
On the one hand, we can follow the crossover from tunneling oscillations (expected for $r\approx {\ell}$) to particle loss  ---realized in the limit of a large width $r\gg {\ell}$.
On the other hand, the corresponding Hamiltonians (\ref{eq.H_tp}) and (\ref{eq.H_sp}) can be numerically exactly diagonalized (see Appendix \ref{sec.app} for details).
The price we pay is a finite observation time, defined by the condition that no reflections from the hard wall boundary of the right well must occur.
We are, however, rewarded with direct access to all spectral quantities of the problem.
This represents a complementary approach to the direct solution of the time-dependent Schr\"odinger equation \cite{LSAMC09,GC11,KB11,LSSAC12}, and yields fundamental insight as to which quasi-continuum states actually support the decay.
This information should also advance the derivation of a microscopic master equation and we shall shortly return to this issue at the very end of this article.

The major theme of our study is under what circumstances and to what extent the bosons tunnel as individual particles or in a correlated manner, i.e., as a  pair.
To facilitate the forthcoming discussion, we visualize the potential landscape of Eq.~(\ref{eq.H_tp}) in the $x$-$y$ plane [see Fig.~\ref{fig.setup}(b)].
The white regions denote the tunneling barrier of height $V_0$, and the dashed line represents the diagonal ($x=y$) at which the contact interaction $U$ arises.
Region (1) corresponds to both particles being in the left well, while regions (2) indicate that one boson is in either well.
Finally, region (3) corresponds to both particles being located in the right well.
In this representation, uncorrelated tunneling of the two bosons would manifest in a transition $(1)\rightarrow(2)\rightarrow(3)$ while correlated tunneling (of a pair of bosons) would correspond to a direct transition $(1)\rightarrow(3)$.
Later, in Eqs.~(\ref{eq.P_1})-(\ref{eq.P_3}), we shall define the associated probabilities with which the two-particle wave function populates the corresponding regions.

As an aside, we point out that the transitions considered above can be regarded as sequential [$(1)\rightarrow(2)\rightarrow(3)$] or non-sequential [$(1)\rightarrow(3)$] double ionization of many-electron atoms.
The continuing interest in the helium problem is driven by the same question as in the present study: 
That is, to identify the role of the inter-electronic interactions and correlations in the ionization process (for recent publications, see, e.g. \cite{FKE06,HMR07,PFNPSB11}).

%=== SINGLE-PARTICLE LOSS ===
%
\section{Single-particle Case}
\label{sec.spp}
As a preliminary step toward the tunneling of two interacting bosons, we recall the single-particle scenario of Hamiltonian (\ref{eq.H_sp}) and investigate the {\it spectral signatures} of the transition from tunneling ($r\approx {\ell}$) to particle loss ($r\gg {\ell}$).
This allows us to introduce ---in the familiar single-particle context---
one of our main spectral tools, the participation ratio (PR) of the initial state in the Hamiltonian's eigenbasis, i.e., the number of eigenstates that mediate the dynamics.
The analysis of the PR constitutes the first part of the present section and will prove beneficial once we turn to the two-particle dynamics.
At the end of this section, we demonstrate the validity of the quasi-continuum approach, by comparing our results for the single-particle spectrum to the time evolution of the initial state and to analytical predictions.
We note that for a single particle in an (asymmetric) polynomial potential, the transition from tunneling oscillations to tunneling decay was studied in Ref. \cite{BK02}.

\subsection{Initial state}
\label{sec.spis}
Throughout this contribution, we focus on the ground state of the isolated left well as the initial state $|\psi(t=0)\rangle$ of the dynamics.
For a {\it single particle} in a box potential of width ${\ell}$, the configuration-space representations of the eigenstates $|\chi_n^{\ell}\rangle$ are sinusoidal \cite{CT1}, i.e.,
\begin{equation}
\chi_n^{\ell}(x) = \sqrt{\frac{2}{{\ell}}} \sin(k^{(sp)}_n x),
\label{eq.sp_eig_state}
\end{equation}
where $k^{(sp)}_n=\pi n/{\ell}$ denotes the single-particle momentum and the corresponding eigenenergies are 
\begin{equation}
\epsilon^{(sp)}_n = (k^{(sp)}_{n})^2=\frac{\pi^2 n^2}{{\ell}^2} \, .
\label{eq.sp_eig_energy}
\end{equation}
Thus, with the numbers from above the ground-state energy $\epsilon_1^{(sp)}=3.79\times 10^{-3}$ is well below the barrier height $V_0=0.1$.
Here and in the following, the super-script $(sp)$ indicates single-particle quantities.
If not stated otherwise, the quantities $k_n^{(sp)}$ and $\epsilon^{(sp)}_n$ pertain to the left well, and hence we drop the index $\ell$.

Since the $|\chi_n^{\ell}\rangle$ are typically not eigenstates of the double-well potential (also referred to as the {\it total system} in the following), we generally observe non-trivial dynamics once we set $|\psi(0)\rangle = |\chi_n^{\ell}\rangle$.
Since numerically exact diagonalization is at the heart of our approach, we obtain the time-evolved state by spectral decomposition.
That is, we expand  $|\psi(0)\rangle$ in terms of the eigenstates $\{|E^{(sp)}_n\rangle \}$ of the total system (\ref{eq.H_sp}),\footnote{\label{foot.c_n}In the two-boson dynamics studied from Sec.~\ref{sec.tp_box} onwards, we expand the (two-body) initial state (\ref{eq.psi_tp}) in terms of the two-particle eigenstates of the total system.}
\begin{equation}
c_n = \langle E^{(sp)}_n |\psi(0)\rangle \, ,
\label{eq.c_n}
\end{equation}
and use the familiar expression
\begin{equation}
|\psi(t)\rangle = \sum_n  c_n e^{-iE^{(sp)}_nt}  |E^{(sp)}_n\rangle \, .
\label{eq.psi_t}
\end{equation}

\subsection{Participation ratio I: From tunneling oscillations to tunneling decay}
\label{sec.sppr}
A robust and intuitive spectral measure to characterize the quantum dynamics is the participation ratio (or participation number)
\begin{equation}
{\rm PR}(|\psi\rangle)= \left[\sum_n |c_n|^4 \right]^{-1} ,
\label{eq.PR}
\end{equation}
which represents the number of eigenstates $|E^{(sp)}_n\rangle$ that significantly contribute to the initial state $|\psi(0)\rangle$.
The participation ratio varies from unity --when $|\psi(0)\rangle$ coincides with one eigenstate-- to $N$, for $|\psi(0)\rangle$ an
equally-weighted superposition of all eigenstates.
Here, $N$ is the dimension of the associated Hilbert space which is always finite in our numerical study (see Appendix \ref{sec.app}).

Yet, as defined in Eq.~(\ref{eq.PR}), the participation ratio does not take into account the spatial density distribution of the $|E^{(sp)}_n\rangle$, i.e., the PR cannot resolve which part of configuration space is populated by the contributing eigenstates.
Such a distinction, however, is highly desirable since it can provide valuable insight into the systems dynamics, as we shall see below.
To achieve this, we employ the probability $P_{\ell}$ to find the particle described by $|\psi(t)\rangle$ in the left
well,\footnote{Note that the probability density within the barrier is negligible. Hence, $P_r$ can be regarded as the probability to find the particle in the right well.}
\begin{equation}
P_{\ell}(|\psi(t)\rangle) =  \int^{\ell}_0 {\rm d}x \  |\psi(t)|^2 \,  \  ; \, P_r=1-P_{\ell}\, .
\label{eq.P_l}
\end{equation}
With the help of (\ref{eq.P_l}) we can define {\it weighted} versions of the PR as follows:
\begin{align}
{\rm PR}_{r,\ell}(|\psi\rangle) = & \left[\frac{\sum_n |c_n|^4 P_{r,\ell}(|E^{(sp)}_n\rangle)^2}{[\sum_n |c_n|^2 P_{r,\ell}(|E^{(sp)}_n\rangle)]^2}\right]^{-1} \nonumber \\
&\nonumber \\
& \times \sum_n |c_n|^2 P_{r,\ell}(|E^{(sp)}_n\rangle) .
\label{eq.PR_weighted}
\end{align}
The numerator in the first term of (\ref{eq.PR_weighted}) weighs each coefficient $|c_n|^2$ with $P_{\ell}(|E^{(sp)}_n\rangle)$ [$P_r(|E^{(sp)}_n\rangle)$], i.e., with the probability  that a particle in the corresponding eigenstate $|E_n\rangle$ is  found in the left (right) well, while the denominator fixes the range of the first term in (\ref{eq.PR_weighted}) between one and $N$.
The last term in Eq.~(\ref{eq.PR_weighted}) represents an additional weighting which we explain in the discussion around Footnote \ref{foot.PR_r}.
Put differently, ${\rm PR}_{l}(|\psi\rangle)$ [${\rm PR}_{r}(|\psi\rangle)$] provides a measure for the number of those eigenstates that participate in the time evolution {\it and} have a fraction in the {\it left} [{\it right}] region of configuration space.

\begin{figure}[t]
\centering
\includegraphics[width=0.98\columnwidth,keepaspectratio]{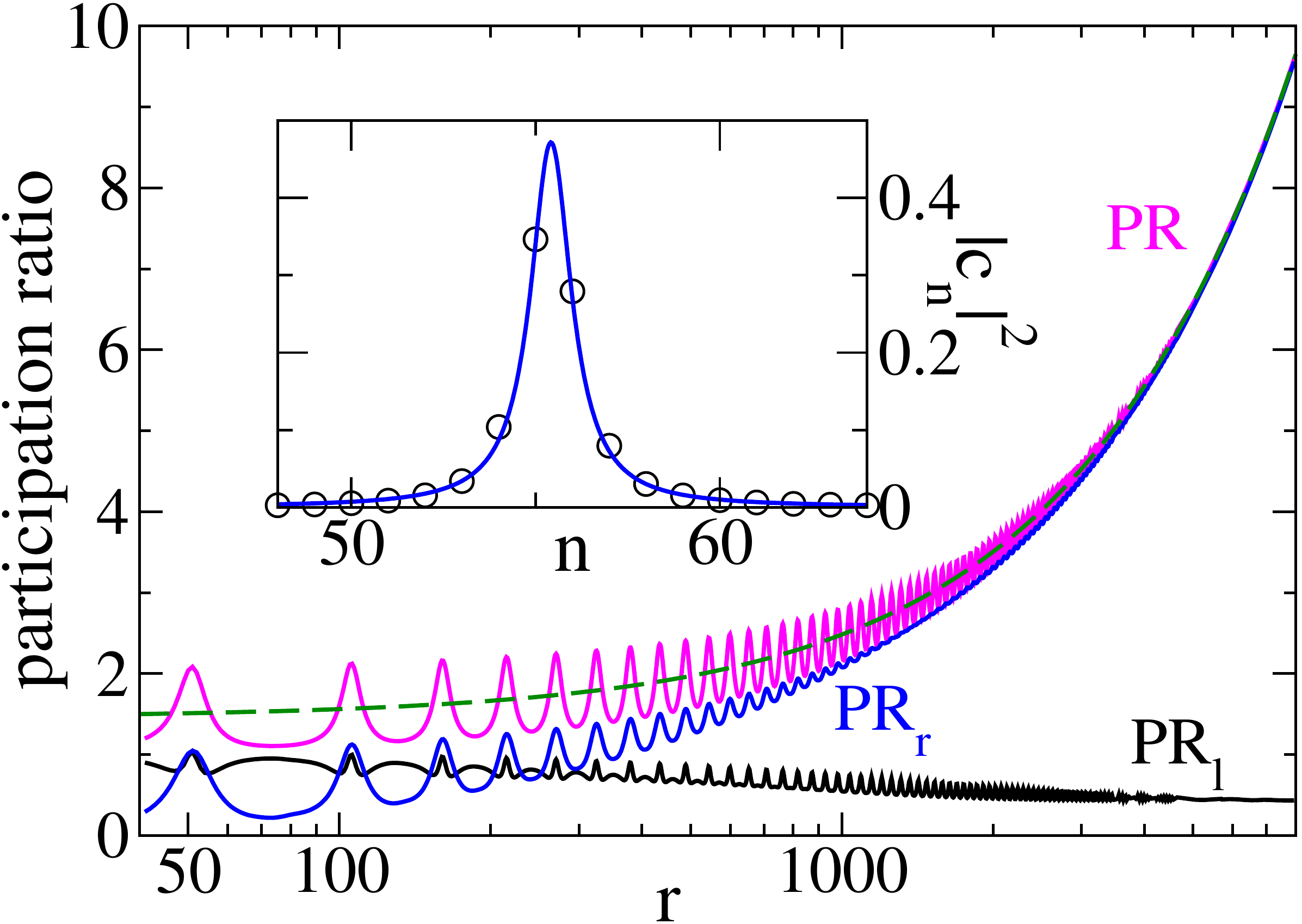}
\caption{ (Color online) Single-particle case:
The participation ratio ${\rm PR}$ (magenta), and the weighted participation ratios ${\rm PR}_{\ell}$ (black) and ${\rm PR}_r$ (blue) for different widths $r$ of the right well [see Eqs.~(\ref{eq.PR}),~(\ref{eq.PR_weighted})].
The dashed green line denotes a linear fit with slope $a=1.02\times 10^{-3}$.
The participation ratios exhibit resonances which increasingly overlap as $r\rightarrow \infty$.
{\it Inset:} The absolute value squared (black circles) $|c_n|^2$ of the expansion coefficients (\ref{eq.c_n}) of the initial state, versus the level index $n$, for a broad right well of width $r=3000$.
The blue solid line represents a Lorentzian fit of width $\Gamma=1.3$.
}
\label{fig.sp_P_Left_PR}
\end{figure}

With these quantities at hand, we now discuss the tunneling dynamics of a particle initially prepared in the ground state  $|\chi_1^{\ell}\rangle$ of the isolated left well, in Fig.~\ref{fig.sp_P_Left_PR}, for variable width $r$ of the right well.

Consider first the familiar case of a symmetric double-well potential ($r=\ell=51$).
The PR takes the value two which can be easily understood by the following argument:
In the symmetric case, the ground states $|\chi_1^{\ell}\rangle$, $|\chi_1^{r}\rangle$ of the isolated left and right well are in resonance.
Thus, the eigenstates $|E^{(sp)}_n\rangle$ of the total system are the symmetric and antisymmetric superpositions of the isolated wells' eigenfunctions, such that the two lowest states read $|E^{(sp)}_{1,2}\rangle\approx(|\chi_1^{\ell}\rangle\pm|\chi_1^r\rangle)/\sqrt{2}$.
Accordingly, the initial state $|\chi_1^{\ell}\rangle$ is a superposition of these two eigenstates and hence the PR equals two.%
\footnote{Throughout the paper we consider the weak coupling regime (see Appendix \ref{app.P_s_w_c}),
otherwise more states could contribute to $|E^{(sp)}_{1,2}\rangle$, even in the case of the symmetric double well.}
As a result, the associated tunneling dynamics displays perfect Rabi oscillations with a period inversely proportional to the energy splitting $E^{(sp)}_2-E^{(sp)}_1$ (not shown).
How does the weighted participation ratio ${\rm PR}_{\ell}$ (black curve) behave?
Said the above it is clear that {\it two} eigenstates contribute to $|\psi(0)\rangle$ which (partially) populate the left well, thus the first term in (\ref{eq.PR_weighted}) equals {\it two}.
Since each of these states amount to half the norm of the initial state and $P_{\ell}(|E^{(sp)}_1\rangle)=P_{\ell}(|E^{(sp)}_2\rangle)=1/2$, the second term equals $1/2$.
In total we have ${\rm PR}_{\ell}=1$ and, due to parity symmetry, also ${\rm PR}_r=1$ (blue curve).

As $r$ increases to, say, $75$, the PR approaches one, indicating that $|\chi_1^{\ell}\rangle$ is approximately an eigenstate of the double-well system.
The physical origin is that at $r=75$, the eigenstates of the isolated right well are maximally detuned from resonance with $|\chi_1^{\ell}\rangle$.
Accordingly, ${\rm PR}_r$ tends to zero\footnote{\label{foot.PR_r} In this case all coefficients $\{|c_n|^2 P_r(|E^{(sp)}_n\rangle)\}$ are approximately zero. However,
due to the normalization via the denominator in Eq.(\ref{eq.PR_weighted}), the PR$_r$ could become very large.
This artificial increase is prevented by the additional weighting with the last term in Eq.(\ref{eq.PR_weighted}).}
and ${\rm PR}_{\ell}\approx1$. 
The associated dynamics is trivial: the particle is persistently trapped in the left well and only small amounts of probability leak out. 

Once $r$ is further increased, the PR exhibits a sequence of equally spaced peaks.
These occur whenever an eigenstate of the isolated right well $|\chi_n^r\rangle$ comes into resonance with the ground state of the isolated left well $|\chi_1^{\ell}\rangle$, i.e., for $r\approx n\ell$, $n=1,2,...$ .
Accordingly, we will refer to the maxima as {\it resonances}.
At these values of $r$, one observes Rabi oscillations while, in between two peaks, the tunneling is suppressed.%
\footnote{We note that a similar situation arises in lattices systems subject to an additional tilt: At appropriately tuned values of the lattice tilt, neighboring sites become resonant, which leads to the resonance-enhanced tunneling (see, e.g. \cite{WMM05}).}

The situation drastically changes once the width $r$ is substantially increased.
Then, the oscillations in the PR are reduced, the PR and the ${\rm PR}_r$ become larger than one, for all $r$, while PR$_{\ell}$ decreases.
In this regime, the widths of the resonances (between eigenstates of the left and the right well) are larger than their respective spacing, which is simply given by the difference of two consecutive eigenenergies $\epsilon_n^{(sp)}$ of the {\it right} well.%
\footnote{\label{foot.res_width}We note that the width of the resonances increases monotonically with $r$.}
In our setup, this transition takes place around $r\approx1000$, that is, $r\ge1000$ denotes the regime of {\it overlapping resonances}, where the density of states in the right well is large enough such that many states contribute to the initial state, and a {\it quasi-continuum} is formed; a prerequisite to observe loss dynamics rather than Rabi oscillations.
Regarding the weighted participation ratios in the limit of large $r$, we observe that PR$_{\ell}$ saturates at a finite value.% 
\footnote{\label{foot.weak_coupl} We note in passing that the asymptotic value of ${\rm PR}_{\ell}$ can be derived by assuming that, within the energy window in which the coefficients $|c_n|^2$ are large, the PR$_{\ell}(|E^{(sp)}_n\rangle)$ follow the same Lorentzian distribution as the coefficients $|c_n|^2$ (see the following section).
We obtain an asymptotic value of $16/40=0.4$ which agrees well with our numerics. 
We numerically confirmed that this assumption on the distribution of the PR$_{\ell}(|E^{(sp)}_n\rangle)$ is indeed justified  for $r\rightarrow\infty$ (see Appendix \ref{app.P_s_w_c}).}
At the same time, ${\rm PR}_r$ approaches PR, since the eigenstates $|E_n^{(sp)}\rangle$ that contribute to the initial state $|\psi(0)\rangle$ have an ever increasing overlap with the right well.

\subsection{Participation ratio II: Tunneling decay rate}
\label{ssec.3c}

Before we turn to the corresponding time evolution, let us formulate the expected particle loss quantitatively.
According to Wigner's theory of decaying systems (see, e.g. \cite{cohen-tann_API}), the coefficients $|c_n|^2$ should follow a Lorentzian distribution of width $\gamma$ in energy space [see also the schematic illustration in Fig.~\ref{fig.LDOS_scheme}a)] leading to an exponential decay behavior in the time domain with rate $\gamma$.\footnote{We note that this holds only  for the initial state $|\psi(0)\rangle = |\chi_n^{\ell}\rangle$ an eigenstate of the isolated left well (see Appendix \ref{app.P_s_w_c}).
Furthermore, we stress that the exponential decay is realized on intermediate time scales, while deviations from it naturally occur for ultra-short and ultra-long times and have even been experimentally observed \cite{WBFMMNSR97,RHM06}.
}
For large enough $r$ and sufficiently weak coupling between the two wells (as in our case), the mean level spacing of the total system $\Delta_n = E^{(sp)}_{n+1} -E^{(sp)}_n$ can be safely assumed as constant within the extension of the Lorentzian.
In this way, we can consider the $|c_n|^2$ as a function of the level index $n$ rather than of the energy $E^{(sp)}_n$, what is much closer to the definition of the PR.
In the inset of  Fig.~\ref{fig.sp_P_Left_PR}, we plot the $|c_n|^2$ versus $n$, together with a Lorentzian fit (solid line) of width $\Gamma=1.3$ for $r=3000$.
The agreement with the theoretical prediction is almost perfect and it remains to connect $\Gamma$ to the decay rate $\gamma$ and, ultimately, to the participation ratio PR.
The first step is a simple proportionality relation:%
\begin{equation}
\gamma= \frac{\Gamma}{\rho^{sp}(\epsilon^{(sp)}_1)}\, ,
\label{eq.gamma_sp}
\end{equation}
where
\begin{equation}
\rho^{sp}(E)=\frac{r}{2\pi}\frac{1}{\sqrt{E}} \,  
\label{eq.rho_sp}
\end{equation}
is the single-particle density of states in the quasi-continuum, evaluated at the energy $E=\epsilon^{(sp)}_1$ of the initial state.
In a second step, we calculate the PR for perfectly Lorentzian distributed expansion coefficients $|c_n|^2$ to be ${\rm PR}(|\psi(0)\rangle) = \pi\Gamma$. Hence, the loss rate becomes
\begin{eqnarray}
\gamma &=& \frac{{\rm PR}(|\psi(0)\rangle)}{\pi \rho^{sp}(\epsilon^{(sp)}_1)} = \frac{2 {\rm PR}(|\psi(0)\rangle)}{r}\sqrt{\epsilon^{(sp)}_1} \, . 
\label{eq.gamma_PR}
\end{eqnarray}
It is physical to assume that for large enough $r$, the rate $\gamma$ of the decay should not depend anymore on the width $r$ of the right well.
We thus conclude from (\ref{eq.gamma_PR}) that the participation ratio grows linearly with $r$, i.e., ${\rm PR}(|\psi(0)\rangle)/r\rightarrow a=const.$ for $r\rightarrow \infty$.
The expected linear increase is confirmed by Fig.\ref{fig.sp_P_Left_PR} where the dashed green line represents a linear fit on the PR.
From the slope $a=1.02\times10^{-3}$ we obtain the loss rate 
\begin{equation}
\gamma(\epsilon^{(sp)}_1) = 2a\sqrt{\epsilon^{(sp)}_1}=1.25\times10^{-4} \, .
\label{eq.gamma_PR_value}
\end{equation}

\subsection{Time evolution and analytical treatment}

\begin{figure}[t]
\centering
\includegraphics[width=0.98\columnwidth,keepaspectratio]{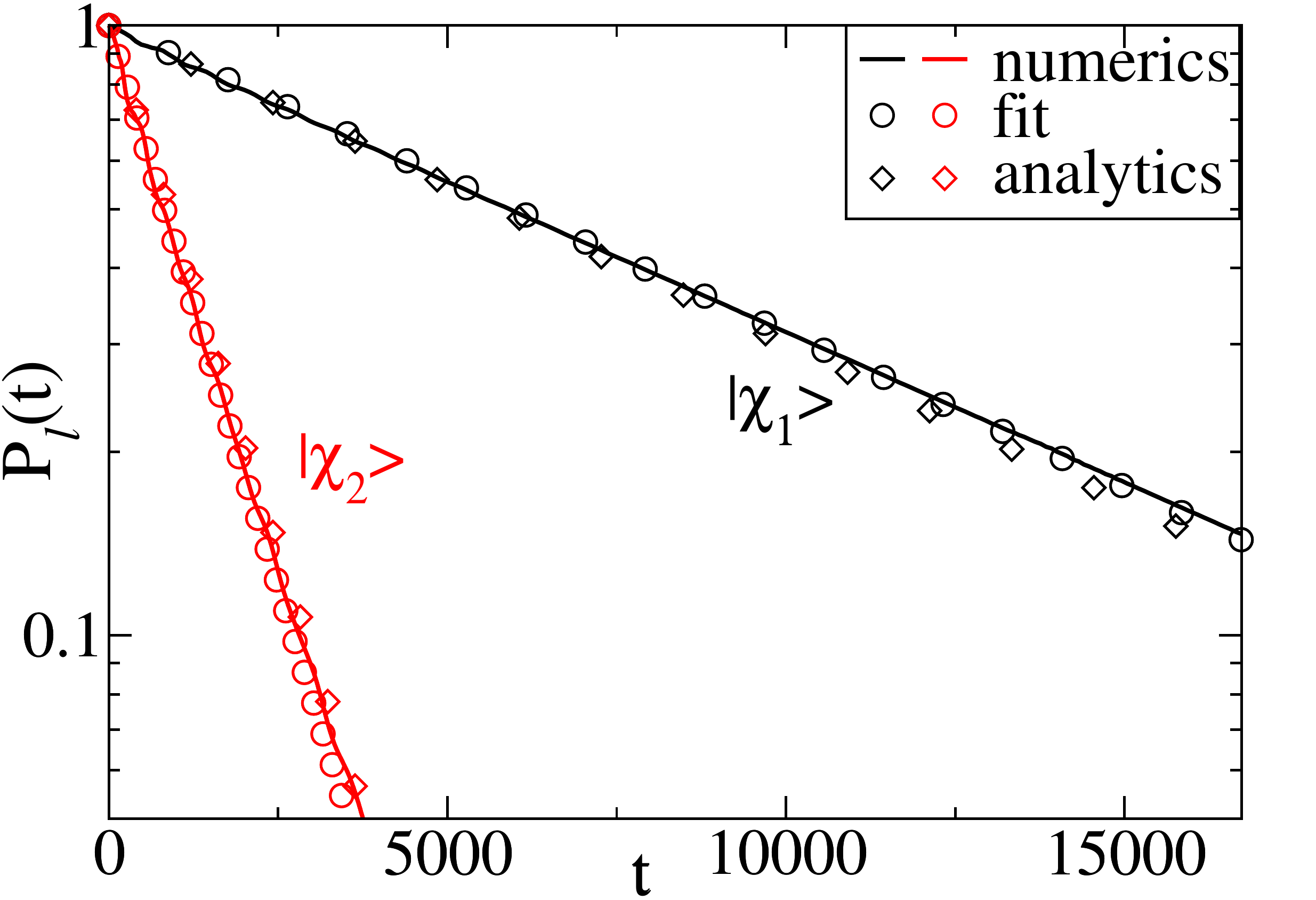}\\
\caption{ (Color online) Single-particle decay:
Semi-logarithmic plot of the probability $P_{\ell}(t)$ %
to find the boson in the left well versus time $t$,  Eq.~(\ref{eq.P_l}).
The black (red) line corresponds to an initial condition defined by the ground state $|\chi_1^{\ell}\rangle$ (first excited state $|\chi_2^{\ell}\rangle$) of the isolated left well (\ref{eq.sp_eig_state}).
Circles denote exponential fits [Eqs.~(\ref{eq.gamma_sp1}),(\ref{eq.gamma_sp2})] while 
diamonds represent the analytical result (see text and Appendix \ref{app.ana_sp}).
}
\label{fig.sp_P_Left}
\end{figure}

We now turn to the particle loss {\it dynamics}, with the probability $P_{\ell}(t)=P_{\ell}(|\psi(t)\rangle)$ that the particle is still located in the left well [see Eq.~(\ref{eq.P_l})], as the relevant observable.
The time evolution is obtained from Eq.~(\ref{eq.psi_t}) by diagonalization of the single-particle Hamiltonian (\ref{eq.H_sp}).
In Fig.~\ref{fig.sp_P_Left}, we plot $P_{\ell}(t)$ versus $t$, for $|\psi(0)\rangle=|\chi_1^{\ell}\rangle$.
We observe an exponential decay, as expected from the Lorentzian behavior in the energy domain.
The dynamical loss rate is obtained by an exponential fit to be 
\begin{equation}
\label{eq.gamma_sp1}
\gamma^{(sp)}_1=1.16\times10^{-4}
\end{equation}
and thus found in good agreement with the spectrally determined rate (\ref{eq.gamma_PR_value}).
In anticipation of the two-particle dynamics discussed further down, we also consider the decay of the first excited state $|\chi_2^{\ell}\rangle$, which will play a role when the inter-particle interaction leads to fermionization of the bosons.
The decay of $|\psi(t=0)\rangle=|\chi_2^{\ell}\rangle$ is plotted as a red line in the same figure, with a fitted decay rate
\begin{equation}
\label{eq.gamma_sp2}
\gamma^{(sp)}_2=8.5\times10^{-4}\,.
\end{equation}
Due to its larger energy, $|\chi_2^{\ell}\rangle$ effectively experiences a lower barrier than $|\chi_1^{\ell}\rangle$, hence $\gamma^{(sp)}_2$ is considerably larger as compared to the rate $\gamma^{(sp)}_1$.

The question naturally arises whether the results derived with the above quasi-continuum approach agree with the actual loss physics to a perfect continuum, associated with a right well that stretches out to infinity.
Therefore, we analytically solve the related scattering problem as a benchmark.
We obtain complex eigenenergies with imaginary parts which be interpreted as loss rates (cf. Appendix \ref{app.ana_sp}).
The induced decay is represented in Fig.~\ref{fig.sp_P_Left} by diamonds, and agrees well with the quasi-continuum results.
We thus conclude that our quasi-continuum approach is indeed the appropriate tool for our study of the temporal behavior on not too long time scales.

In the familiar single-particle case, we thus successfully used the participation ratio to observe the transition from tunneling oscillations (isolated resonances) to tunneling decay (strongly overlapping resonances).
In the next sections, we will generalize this concept to the case of two interacting bosons.

%=== INITIAL STATE ===
%
\section{Two-particle case: Initial state and physical observables}
\label{sec.tp_box}
\subsection{Ground state of two bosons in a box potential as initial state}
\begin{figure*}[t]
\centering
\includegraphics[width=0.63\columnwidth,keepaspectratio]{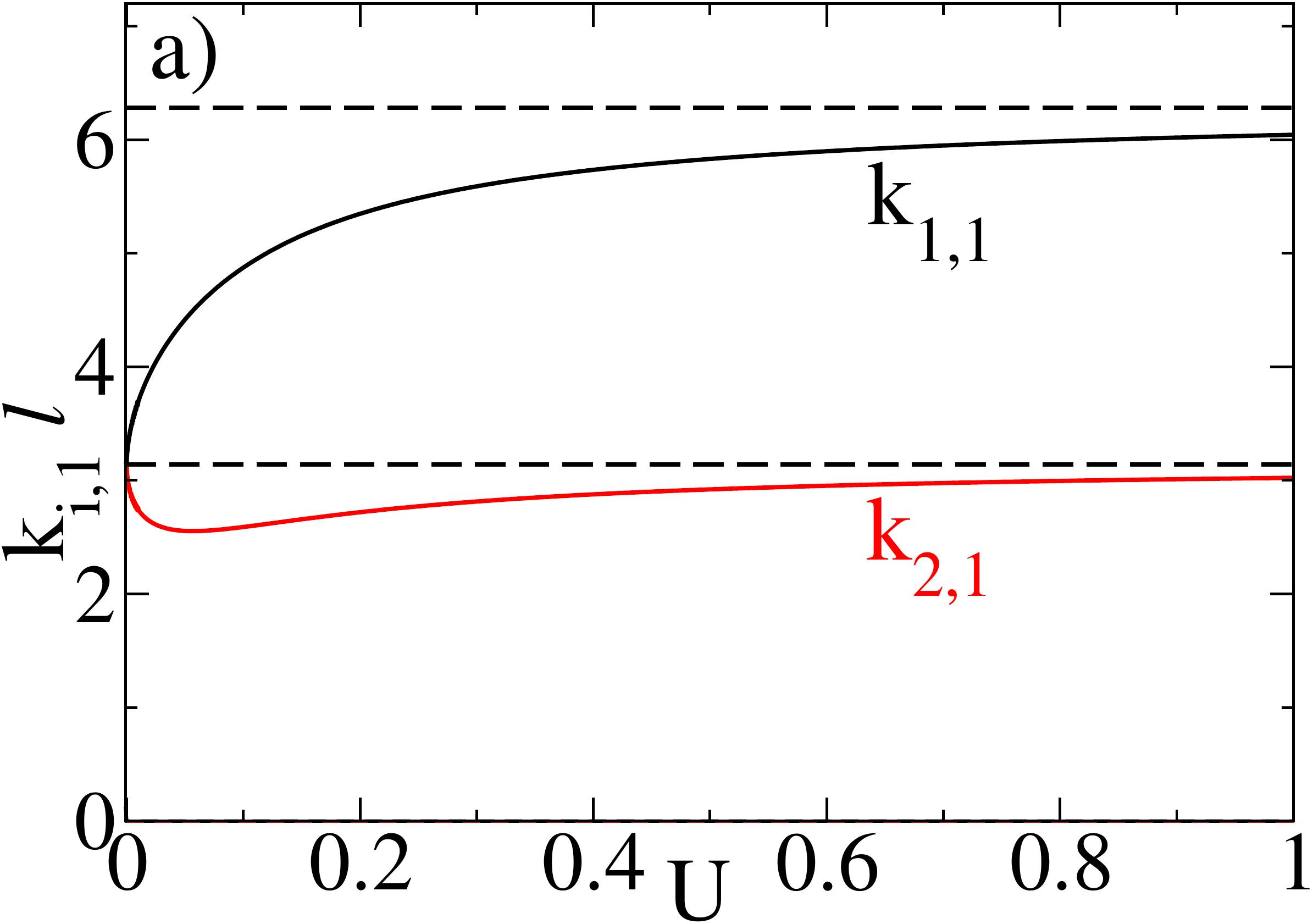}\hfill
\includegraphics[width=0.64\columnwidth,keepaspectratio]{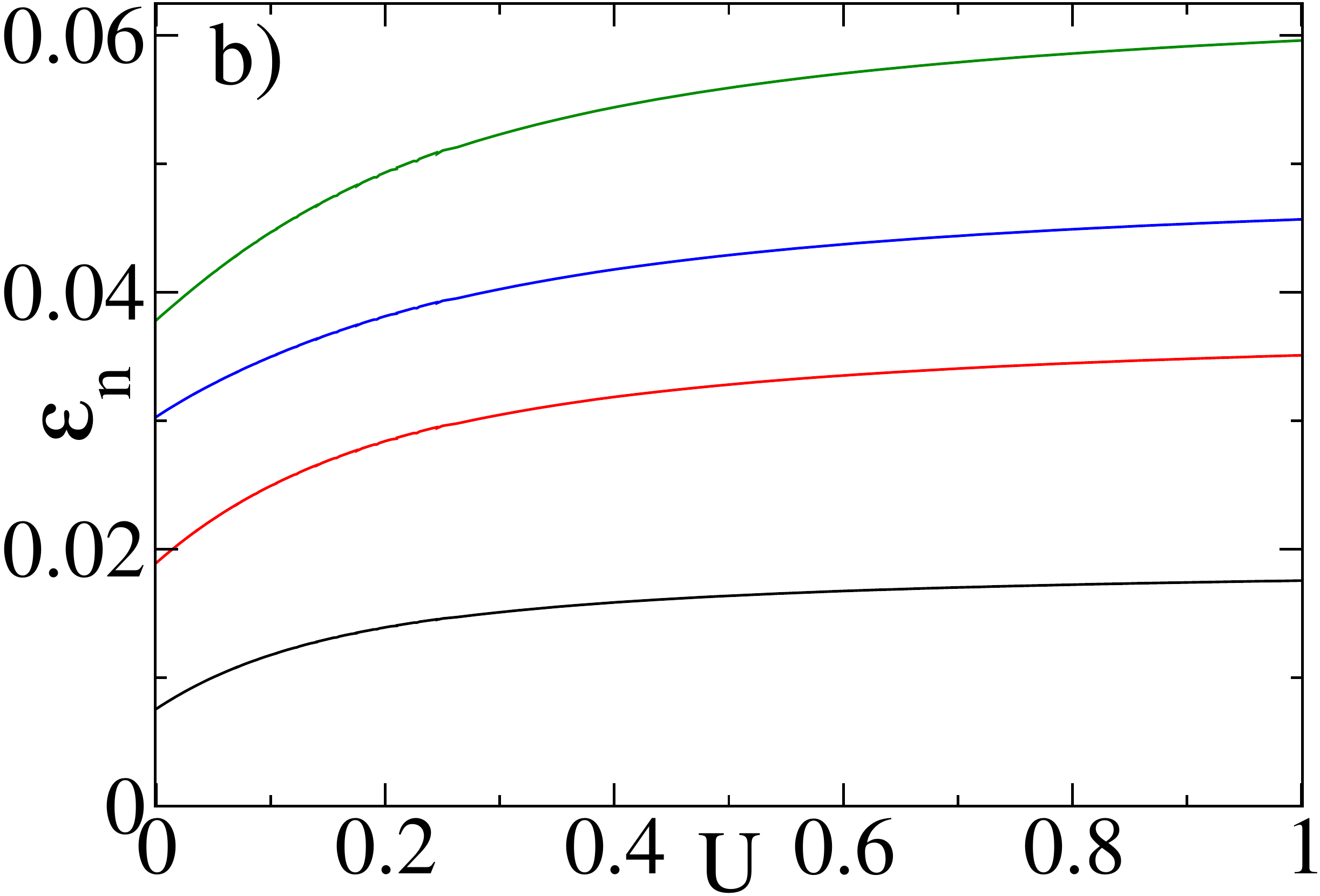}\hfill
\includegraphics[width=0.635\columnwidth,keepaspectratio]{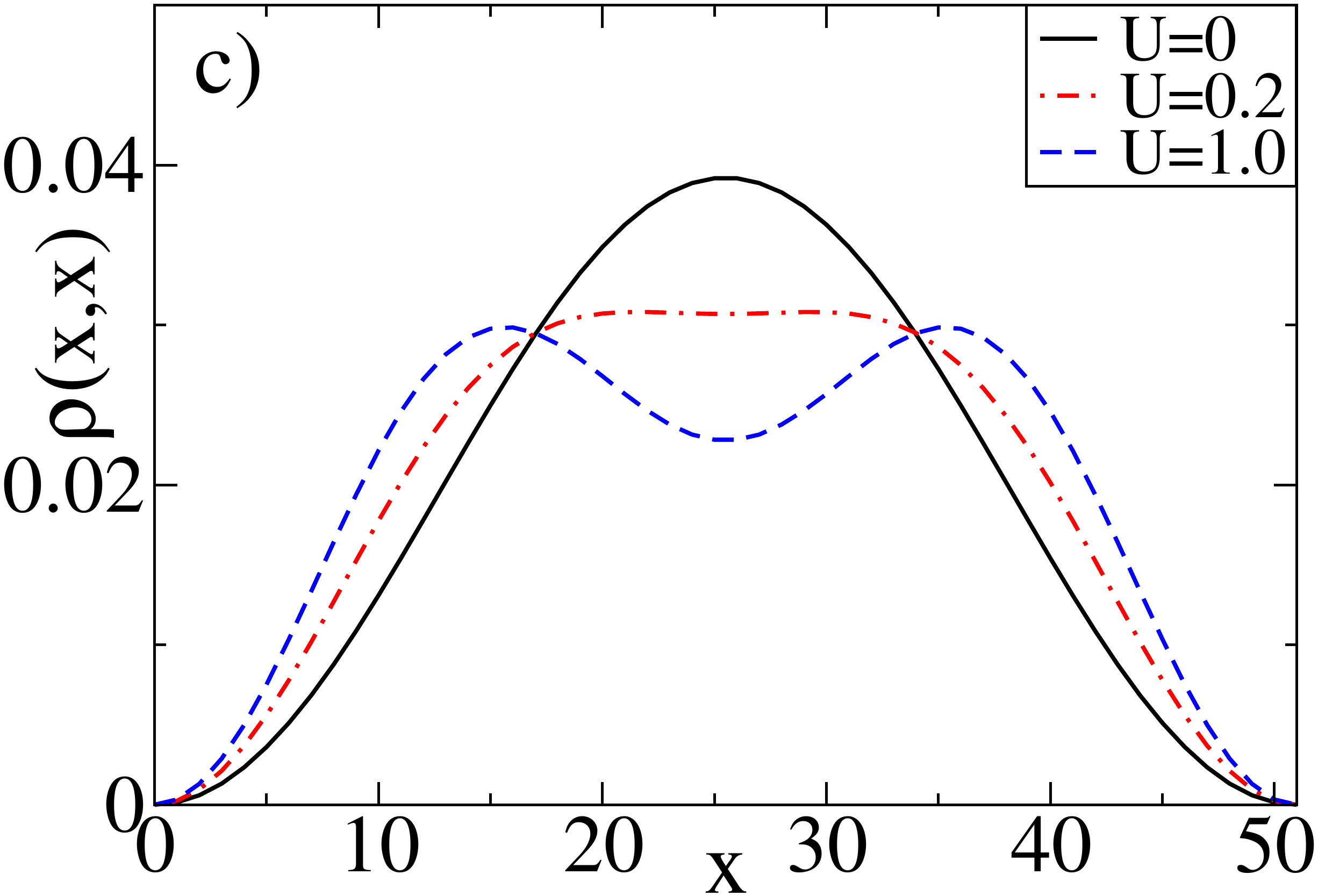}\\ \vspace{0.2cm}
\caption{
Various observables of two interacting bosons in a single-well potential of width $\ell=51$:
(a) The ground state momenta $k_{1,1}$ (black) and $k_{2,1}$ (red) [see (\ref{eq.k_1_2})], in units of the inverse width $1/\ell$ of the well, versus the inter-particle interaction strength $U$.
The dashed horizontal lines denote the asymptotic values $k\ell= \pi$, and $2\pi$, respectively.
(b) Lowest two-particle energy levels versus $U$, obtained from the exact numerical diagonalization of Eq.~(\ref{eq.H_tp}).
(c) Diagonal part $\rho_{red}(x,x)$ of the reduced single-particle density matrix (\ref{eq.rho_red}) of the respective two-particle ground state, versus position $x$, for various values of $U$. The repulsive interaction reduces the probability of finding a boson at the center of the trap, at $x=\ell/2$.
}
\label{fig.tp_box_spec_momenta}
\end{figure*}
We now turn to the case of two interacting particles.
As a first step, we define the initial state of the system, i.e., an eigenstate of two bosons in the isolated left well of width $\ell=51$. 
The latter can be obtained analytically by means of a Bethe ansatz \cite{Bethe31} from the corresponding Hamiltonian (\ref{eq.H_tp}) with $r=0=b$.
More precisely, the $n$-th eigenfunction $\psi_{k_{1,n}k_{2,n}}(x,y)$ is completely characterized by two wave vectors $k_{1,n}$ and $k_{2,n}$ and reads, for $y\le x$ \cite{LL63,Gaudin71}:
\begin{widetext}
\begin{eqnarray}
\begin{split}
\psi_{k_{1,n}k_{2,n}}(x,y) &\propto\,\, [ A_1\exp{(ik_{1,n}x)} - A_2 \exp(-ik_{1,n}x)] \sin(k_{2,n}y)  \\
					&\,\,\,+ [ A_3\exp{(ik_{2,n}x)} - A_4 \exp(-ik_{2,n}x)] \sin(k_{1,n}y)  \, .
\label{eq.psi_tp}
\end{split}
\end{eqnarray}
\end{widetext}
The amplitudes $A_i(k_{1,n},k_{2,n})$ are functions of the two-particle wave vectors $k_{i,n}$ which, in turn, depend on the inter-particle interaction strength $U$ and represent the solutions of the coupled equations:
\begin{widetext}
\begin{eqnarray}
\begin{split}
k_{1,n}  \ell &= n_1\pi + \arctan\left[\frac{U}{k_{1,n} -k_{2,n}}\right]  + \arctan\left[\frac{U}{k_{1,n} + k_{2,n}}\right] \,,\\
k_{2,n}  \ell &= n_2\pi - \arctan\left[\frac{U}{k_{1,n} -k_{2,n}}\right]  + \arctan\left[\frac{U}{k_{1,n} + k_{2,n}}\right] \, . 
\label{eq.k_1_2}
\end{split}
\end{eqnarray}
\end{widetext}
For $U=0$, the equations decouple and we immediately recover the single-particle result in which the two positive integers $n_{1,2}$
are simply the quantum numbers of a single particle in a box of size $\ell$. 
In the general case $U\neq0$, the $k_{i,n}$  can be regarded as the momenta of two bosons with corresponding two-particle energy
\begin{equation}
\label{eq.E_tp}
\epsilon^{}_n = k^2_{1,n} + k^2_{2,n}\,.
\end{equation}
As in the single-particle case, the quantities $k_{i,n}$ and $\epsilon_n$ pertain to the left well (if not stated otherwise).

Since we are dealing with ultracold atoms, we are predominantly interested in the initial state being the ground state [i.e., $n_1=n_2=1$ in Eq.~(\ref{eq.k_1_2})].
In the remainder of this subsection, we recall its properties as a function of the interaction strength $U$. 
We characterize the ground state by the momenta $k_{i,1}$ and the diagonal part of the associated single-particle reduced density-matrix %
\footnote{In this section, $\psi(x,y,t)=\psi_{k_{1,1}k_{2,1}}(x,y)$ while we consider time-dependent wave functions further on.} 
\begin{equation}
\rho_{red}(x,x',t) = \int {\rm d}y\ \psi^{*}(x,y,t)\psi^{}(x',y,t) \, .
\label{eq.rho_red}
\end{equation}
The latter quantity determines all single-particle quantities such as, for example, the number of bosons in a specific region of configuration space [compare Eq.~(\ref{eq.N_L})].

In Fig.~\ref{fig.tp_box_spec_momenta}a), we plot the momenta $k_{1,1}$ and $k_{2,1}$ of the ground state, in units of the inverse width $1/\ell$, versus the inter-particle interaction strength $U$. 
For $U=0$, the bosons are independent of each other, and both momenta take the single-particle value $k_{i,1}\ell=\pi$.
Accordingly, the two-particle ground-state wave function at $U=0$ is a product of single-particle wave functions,
\begin{equation}
\psi^{(U=0)}_{k_{1,1}k_{2,1}}(x,y) = \chi_1(x) \ \chi_1(y) \propto \sin(k^{(sp)}_1x)\ \sin(k^{(sp)}_1 y) \, ,					
\label{eq.psi_u=0}
\end{equation}
with a two-particle ground-state energy $\epsilon^{}_1(U=0) = 2\epsilon^{(sp)}_1 $ given by twice the single-particle ground state energy (\ref{eq.sp_eig_energy}).
Since both bosons occupy the same single-particle state, also the diagonal of the corresponding reduced single-particle density matrix [panel c)] is perfectly sinusoidal.

As $U$ increases, the particles repel each other and the two-particle energy grows, until it saturates at the energy of the first excited state $\psi^{(U=0)}_{k_{1,2}k_{2,2}}(x,y)$:
 this is evident from panel (b) where we plot the parametric evolution of the lowest-lying energy levels $\epsilon^{}_n$, obtained from the exact diagonalization of Eq.~(\ref{eq.H_tp}), with $r=0=b$ and $l=51$.\footnote{These numerical results perfectly agree with the analytical result (\ref{eq.k_1_2}).}
In the reduced single-particle density matrix, this repulsion between the bosons is manifest in a reduced probability to detect a boson at the center of the box, i.e. at $x=\ell/2$, see panel c).

This observation can immediately be understood in terms of the wave vectors: One finds that the (larger) momentum $k_{1,1}$ grows with $U$, and asymptotically reaches $k_{1,1}\ell=2\pi$, while $k_{2,1}$ first slightly decreases and asymptotically approaches $k_{1,1}\ell=\pi$ \cite{KB11}.
The indistinguishability of the particles aside, one may imagine that for $U\rightarrow\infty$ one boson is in the single-particle ground state $|\chi^\ell_1\rangle$, while the other one is in the single-particle first excited state $|\chi^\ell_2\rangle$, a situation also referred to as the fermionization or Tonks-Girardeau limit \cite{Girardeau60} of a repulsively interacting Bose gas.
In this limit, the wave function exactly vanishes on the diagonal where the (contact) interaction arises and is given by
\begin{align}
\begin{split}
\psi^{(U=\infty)}_{k_{1,1}k_{2,1}}(x,y) \propto \frac{1}{\sqrt{2}} &\left[ \sin(k^{(sp)}_2y)\ \sin(k^{(sp)}_1 x) - \right. \\
&\left.\sin(k^{(sp)}_2x)\ \sin(k^{(sp)}_1 y) \right] \, ,
\end{split}
\label{eq.psi_u_infty}
\end{align}
for $y\le x$, while its energy $\epsilon^{}_1(U=\infty)=\epsilon^{(sp)}_1 +\epsilon^{(sp)}_2$ is the sum of the single-particle ground and first-excited state energies. Of course, $\psi_{k_{1,1}k_{2,1}}(x,y)$ always retains its bosonic symmetry with respect to exchange of $x$ and $y$.
Thus,  the ground state wave function of the (interacting) bosonic and of the corresponding (interaction-free) fermionic system coincide in modulus, but have opposite sign for $x>y$.

\subsection{Physical observables}
\label{sec.tp_phys_obs}
In our analysis of tunneling decay, we initially prepare the ground state $|\psi(t=0)\rangle = |\psi_{k_{1,1}k_{2,1}}\rangle$, Eq.~(\ref{eq.psi_tp}), of two bosons in the isolated left well ($V_0=\infty$) at a given interaction strength $U$, and monitor the ensuing dynamics after the potential barrier is instantaneously set to $V_0=0.1$.
To distinguish uncorrelated single-particle tunneling from correlated two-particle tunneling, we define --according to Fig.~\ref{fig.setup}b)-- the following observables:
The probability $P_1(|\psi(t)\rangle)$ that both particles are in the left well [i.e. in region (1)], the probability $P_2(|\psi(t)\rangle)$ that one boson is in each of the wells [i.e. in region (2)],
\footnote{As is apparent from Fig.\ref{fig.setup}b), region (2) consists of two sub-regions.
Justified by the bosonic symmetry of the wave function $\psi(x,y,t)$, we integrate solely over one of them, hence in Eq.~(\ref{eq.P_2}), the factor of two appears.}
and the probability $P_3(|\psi(t)\rangle)$ that both bosons are located in the right well [i.e. in region (3)],
\begin{eqnarray}
P_1(|\psi(t)\rangle) &=& \int_I {\rm d}x \ \int_I {\rm d}y\ |\psi(x,y,t)|^2 \label{eq.P_1} \,,\\
P_2(|\psi(t)\rangle) &=&2 \int_I {\rm d}x \ \int_{III} {\rm d}y\ |\psi(x,y,t)|^2 \label{eq.P_2}\,, \\
P_3(|\psi(t)\rangle) &=& \int_{III} {\rm d}x \ \int_{III} {\rm d}y\ |\psi(x,y,t)|^2 \, .
\label{eq.P_3}
\end{eqnarray}
For example, uncorrelated tunneling of the two bosons would manifest in a transition $(1)\rightarrow(2)\rightarrow(3)$, i.e., first $P_2$ would rise and then $P_3$, while correlated pair tunneling would correspond to a direct transition $(1)\rightarrow(3)$, i.e., $P_2$ would remain zero during the time evolution.
Due to the normalization of $\psi(x,y,t)$, the $P_i$ sum up to unity.

These truly two-particle quantities are complemented by the experimentally easily measurable \cite{ZSLWRBJ12} number of bosons in the left well,
\begin{eqnarray}
N_{\ell}(t) = \int_0^{\ell} {\rm d}x \ \ \rho_{red}(x,x,t) = P_1(t) + \frac{1}{2} P_2(t) \, ,
\label{eq.N_L}
\end{eqnarray}
which is normalized to a maximal value of one (i.e., $N_\ell=1$ corresponds to two bosons in the left well) and where $\rho_{red}(x,x,t)$ is the diagonal part of the reduced density matrix (\ref{eq.rho_red}).
With this normalization, $N_{\ell}(t)$ is the two-particle counterpart of the single-particle quantity $P_{\ell}(t)$ [see Eq.~(\ref{eq.P_l})].
Further measures based on the reduced single-particle density matrix are introduced in Sec. \ref{sec.tp_dyn_reduced_quant}

%=== INTERACTING PARTICLE LOSS SPECTRUM===
%
\section{Decay of two interacting particles I: Spectral properties}
\label{sec.tp_asdw_spec}
We now turn to the main object of our present study, the tunneling decay of two interacting bosons.
We in particular wish to clarify which continuum states actually support the decay, and
under what conditions correlated tunneling of a boson pair is observed, rather than independent tunneling of the particles.
Taking the symmetric double-well as starting point, we gradually increase the width $r$ of the right well and monitor the transition from tunneling oscillations to the regime of particle loss ($r\gg \ell$), much as in the single-particle case studied in Sec.~\ref{sec.spp}.
In what follows, we will first analyze characteristic spectral properties and then, in Sec.~\ref{sec.tp_asdw_dyn},  turn to the ensuing dynamics for vanishing and repulsive interaction strengths $U$.

\begin{figure}[t]
\centering
\includegraphics[width=0.95\columnwidth,keepaspectratio]{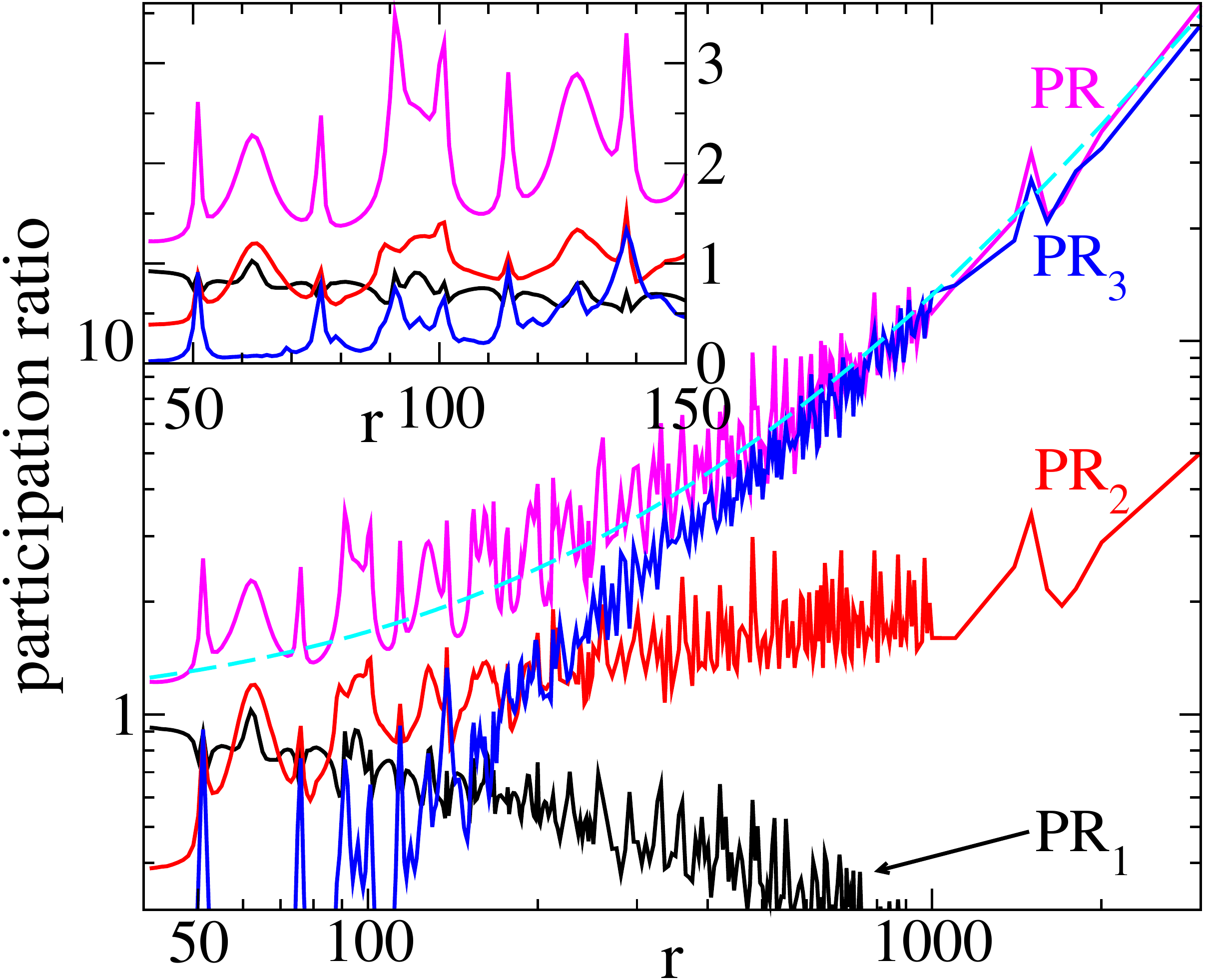}
\caption{(Color online) Two-particle case:
The participation ratio ${\rm PR}$ [magenta, see Eq.~(\ref{eq.PR})] of the initially left localized state $|\psi(0)\rangle$ (\ref{eq.psi_tp}), as well as its weighted versions (see beginning of Sec.~\ref{subsec.PR_tp} for definition) ${\rm PR}_1$ (black), ${\rm PR}_2$ (red), and ${\rm PR}_3$ (blue), for different widths $r$ of the right well, $U = 0.2$, $l=51$, $b=2$, and $V_0=0.1$.
The dashed cyan curve represents a fit with a second-order polynomial $y=ax^2+bx+c$, see the discussion of Eq.~(\ref{eq.tp_N}).
Note that the width $r$ is sampled at larger intervals for $r>1000$.
The inset shows a magnification for $r<150$.
At these small values of $r$, one can clearly distinguish peaks of different width in the PR, which we refer to as {\it resonances}.
Since broad resonances also appear in PR$_2$, we associate these with single-particle tunneling.
Narrow resonances dominate the PR$_3$ and are thus identified with two-particle tunneling, see text.
As the width $r$ increases, the resonances overlap, which renders exclusive one- or two-particle tunneling unlikely.
}
\label{fig.PR_i}
\end{figure}

\subsection{Participation ratio}
\label{subsec.PR_tp}
In complete analogy with Sec.\ref{sec.spp}, we  analyze the participation ratio PR (and its weighted versions PR$_i$) of the initial state.
The definitions (\ref{eq.PR}) of the PR and (\ref{eq.PR_weighted}) of the PR$_i$ directly carry over to the two-particle case.
The only difference is that, in Eq.~(\ref{eq.PR_weighted}), the index is not $i=r,l$ but $i=1,2,3$, since the configuration space is no longer partitioned into {\it left/right}, but consists of the three distinct regions (1), (2), and (3), see Fig.\ref{fig.setup}b).  
Accordingly, we replace in Eq.~(\ref{eq.PR_weighted}) the probabilities $P_{r,\ell}$ with $P_{1,2,3}$ [see Eqs.~(\ref{eq.P_1}-\ref{eq.P_3})] and the $\{c_n\}$ now represent the expansion coefficients of the initial state $|\psi(0)\rangle$ in the two-particle energy eigenbasis $\{|E_n\rangle \}$ of the double-well potential, obtained from diagonalization of Hamiltonian (\ref{eq.H_tp}).
As argued in our discussion of Eq.~(\ref{eq.PR_weighted}), the ${\rm PR}_i$ yield the number of eigenstates that mediate the time evolution, weighted with their overlap with the respective regions (1), (2), or (3).

It will prove beneficial to identify the components of the eigenfunctions $|E_n\rangle$ in terms of the eigenstates of the uncoupled double well ($V_0=\infty$). 
To ease the discussion we label the latter as 
\begin{equation}
|a_i,b_j\rangle \, .
\label{eq.tp_state_fock}
\end{equation}
In this notation, $a$ $(b)$ bosons occupy the $i$-th ($j$-th) eigenstate of the isolated left (right) well, at given interaction strength $U$.
The initial state, for example, corresponds to $|\psi(t=0)\rangle=|2_1,0\rangle$, where the component $|2_1\rangle$ is given by the two-particle ground state (\ref{eq.psi_tp}).
It is important to realize that in a (un-)coupled double-well potential, only states of the types $|2_i,0\rangle$ and $|0,2_i\rangle$ ---where both bosons are in one well--- depend on the interaction $U$, while the energy of a state $|1_i,1_j\rangle$ is insensitive to the short-ranged contact interaction.
Hence, the energy of the latter is simply the sum of the single-particle energies
$\epsilon_i^{(sp)}+\epsilon_j^{(sp)}$,\footnote{Here, $\epsilon_j^{(sp)}$ is given by Eq.~(\ref{eq.sp_eig_energy}), but with $\ell$ replaced by $r$.}
and each of the components $|1_i\rangle$ ($|1_j\rangle$) corresponds to the single-particle wave function $|\chi^\ell_i\rangle$ ($|\chi^r_j\rangle)$, see Sec.~\ref{sec.spis}.

All four participation ratios PR and PR$_i$ are plotted in Fig.~\ref{fig.PR_i} versus the width $r$ of the right well, for the initially left-localized state $|\psi(0)\rangle$ (see Sec.~\ref{sec.tp_box}) and an exemplary interaction strength
$U=0.2$.\footnote{As long as the interaction is repulsive $U>0$, one merely observes quantitative differences in the PR, concerning, e.g., the position and height of its peaks. %
For the special case of non-interacting particles ($U=0$), the PR shows equally spaced resonances of similar width as in the single-particle case, but now the three states $|2_1,0\rangle$, $|0,2_i\rangle$ and $|1_i,1_j\rangle$ participate.}

To familiarize ourselves with the quantities at hand, we first consider intermediate widths $r<150$ of the right well (see inset), where one can resolve single resonances in the PR.
As in the single-particle case, they indicate that the initial state $|\psi(0)\rangle$ is in resonance with an eigenstate of the uncoupled system.
Accordingly, we expect pronounced tunneling only for the resonant configurations (see discussion in Sec.~\ref{sec.sppr}).

Yet, the appearance of markedly distinct resonance widths is striking:
Broad resonances occur in PR at $r\approx65, 95, 130, ...$ and are accompanied by broad maxima of ${\rm PR}_2$.
In turn, the narrow resonances in PR, found at $r\approx51,75,90,98,113...$, go along with marked peaks in ${\rm PR}_3$.
The weighted participation ratios unequivocally show that broad resonances  result from the initial state $|\psi(0)\rangle$ being resonant with a state of type  $|1_1,1_i\rangle$ [region(2)] while narrow resonances indicate that the resonance condition with a state of type $|0,2_i\rangle$ [region(3)] is met.

The emergence of different widths constitutes a major observation and a fundamental difference  with respect to the single-particle case (compare Fig.~\ref{fig.sp_P_Left_PR}), in which a sequence of equally spaced resonances of monotonically increasing width was observed (see also Footnote~\ref{foot.res_width}).
There, we had to consider only resonances between eigenstates of the isolated left and right wells.
In the notation of (\ref{eq.tp_state_fock}) this amounts to resonances between $|1_1,0\rangle$ and $|0,1_i\rangle$ states.
In the two-particle case, not only the resonances between $|2_1,0\rangle$ and $|0,2_i\rangle$ but also those between $|2_1,0\rangle$ and $|1_i,1_j\rangle$  states have to be taken into account.

Given that, the origin of the different widths is explained by a simple argument:
The resonance width reflects the coupling strength between the participating states.
Tunneling of a single particle is a {\it first-order} process of large coupling strength (and broad resonances), while the (two-particle) pair-wise tunneling observed for repulsive interactions represents a {\it second-order} single-particle process of substantially smaller coupling and ---correspondingly--- narrow resonances.~\footnote{This is consistent with the observation that the pair-wise tunneling (in the symmetric double well) displays much larger oscillation periods (determined by the coupling strength between the wells) than the corresponding single-particle tunneling processes \cite{ZMS08}.}
More precisely, we have found that the two states $|2_1,0\rangle$ and $|0,2_i\rangle$ are primarily coupled via off-resonant states of the type $|1_1,1_i \rangle$.
This is corroborated by the observation that, for configurations for which the PR displays narrow resonances, PR$_2$ also assumes values on the order of $1$, i.e., states of the type $|1_1,1_i \rangle $ are also involved in the time evolution.
In other words, the observed pair-wise tunneling does not correspond to a {\it direct} tunneling process from region (1) to region (3), but rather involves an intermediate visit to region (2), although the corresponding probability $P_2(t)$ is small for all times.~\footnote{For the symmetric double-well, we shall report on this in detail elsewhere.}
A similar situation is found in the so-called chaos-assisted tunneling \cite{TU94}, where tunneling between regular parts of the underlying phase space is mediated by quantum states which populate the chaotic part of phase space.

That is, we can tell by the width of the resonance whether single-particle or pair-wise tunneling prevails, and we confirmed this prediction
by the direct time evolution of the initial state, for various configurations.
The reader is warned though that, in order to observe the (second-order) pair-wise tunneling, states of the type $|1_i,1_j\rangle$ should be far off-resonant with the initial state.
Otherwise, the dynamics is dominated by the (first-order) single-particle tunneling which, due to the larger coupling, is a faster process.

For large widths $r>150$, the overlap between consecutive resonances grows and the peaks smear out.
Among the weighted quantities, PR$_3$ exhibits the strongest increase and quickly approaches PR, while PR$_1$ drops ---much as in the single-particle case.
PR$_2$ increases on average and, most importantly, does not drop below a level of 1.
This is a first indication that {\it pure} two-particle loss is unlikely to be observed in the dynamical evolution, since the contributing eigenstates have non-vanishing overlap with {\it all} three regions of configuration space.
That is, $|\psi(0)\rangle$ generally is in resonance with  $|0,2_i\rangle$- {\it and} $|1_1,1_i\rangle$-type states.
{\it A priori}, this does not exclude simultaneous one- and two-particle tunneling.
Given our above conjecture that single-particle coupling is much stronger than two-particle coupling, the former is likely to dominate the tunneling process.
In the next section we elaborate on this point, and further develop our spectral picture of the decay of interacting bosons.

\subsection{Density of states}
\label{subsubsection.DOS_U>0}

\begin{figure}[t!]
\centering
\includegraphics[width=1\columnwidth,keepaspectratio]{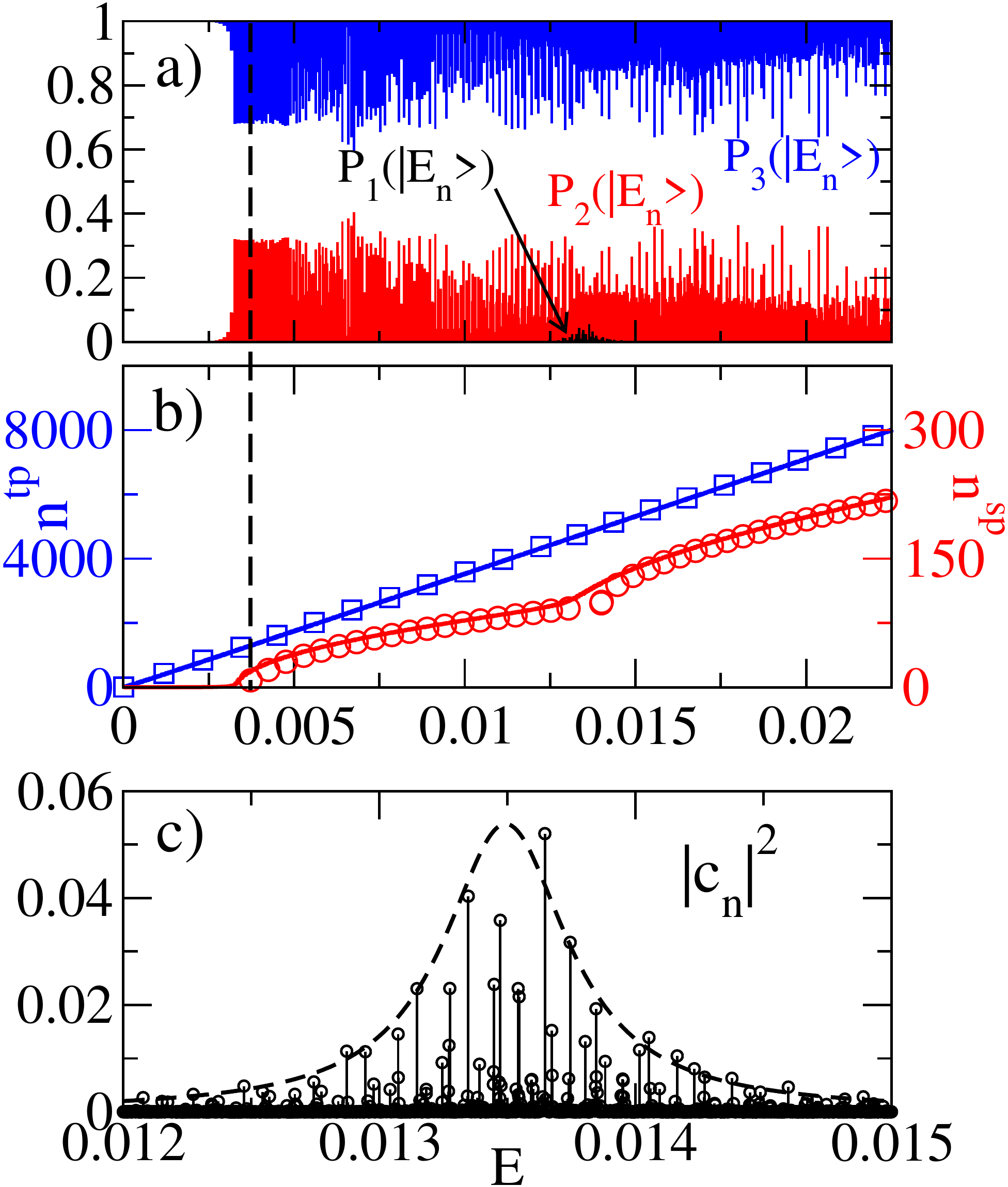}
\caption{(Color online)
{\it Top:} Spatial projections $P_i(|E_n\rangle)$ [see Eqs.~(\ref{eq.P_1})-(\ref{eq.P_3})] of the eigenstates $|E_n\rangle$ of the asymmetric double well versus the corresponding energies $E_n$, for $r=3000$ and otherwise same parameters as in Fig.~\ref{fig.PR_i}.
Up to approximately the single-particle ground-state energy of the left well, $\epsilon^{(sp)}_1=0.0038$ (vertical dashed line), all particles are dominantly localized in the right well [region(3)], since $P_3(|E_n\rangle)=1$.
Above this energy, $P_2(|E_n\rangle)> 0$ indicates eigenstates which (partially) populate region (2), with one boson in each well.
First peaks in $P_1(|E_n\rangle)$ ---corresponding to eigenstates with a doubly occupied left well--- appear at about the energy of the two-particle ground state in the isolated left well $\epsilon_1^{}=0.0139$, see Fig.~\ref{fig.tp_box_spec_momenta}b). 
{\it Middle:} The integrated single- (red) and two-particle (blue) density of states, for the same parameters as in a).
Please mind the two $y$-axis scales.
Solid lines represent the numerical results (\ref{eq.sp_N_num}) and (\ref{eq.tp_N_num}), while symbols correspond to the theoretical predictions (\ref{eq.sp_N}) and (\ref{eq.tp_N}).
We note that, for our specific choice of parameters, the peak in $P_1$ of panel a) (given by the {\it two-particle} energy $\epsilon_1$) accidentally coincides with the kink in $n^{(sp)}$ which results from the first excited {\it single-particle} state with energy $\epsilon^{(sp)}_2=0.0152$.
{\it Bottom:} Only a fraction of the expansion coefficients $|c_n|^2$ 
of the initial state in the double-well's two-particle eigenstates (see Eq.~(\ref{eq.c_n}) and Footnote \ref{foot.c_n}) takes non-zero values in the relevant energy window around $E=\epsilon_1=0.0135$.
The dashed line represents a Lorentzian curve of width $\gamma=5.9\times 10^{-4}$ 
(\ref{eq.gamma_tp_lorentz}). 
See text for details on the extraction of $\gamma$.
}
\label{fig.INT_DOS}
\end{figure}

Let us now attempt to quantify the density of those states (DOS) in the quasi-continuum which respectively support single- and two-particle tunneling.
This is essential if we want to convert the participation ratio into a decay rate as in Eq.~(\ref{eq.gamma_PR}), derived above in Sec.~\ref{sec.sppr} for the single-particle case. 
At the same time, the DOS is of fundamental interest since it represents a key ingredient in the physics of decaying systems, see, e.g., Ref.~\cite{cohen-tann_API}.
To this end, we elaborate on the projections $P_i(|E_n\rangle)$ of the eigenstates [obtained by inserting the $|E_n\rangle$ in Eqs.~(\ref{eq.P_1}-\ref{eq.P_3})] on the three regions of configuration space as defined by Fig.~\ref{fig.setup}b).
For the exemplary case of $U=0.2$ (used before) and a broad right well $r=3000$ ($\ell=51$), the corresponding quantities $P_i(|E_n\rangle)$ are plotted versus the {\it total} energy $E_n$, in Fig.~\ref{fig.INT_DOS}a).

Up to a threshold energy of $E=0.0033$, we find that $P_3(|E_n\rangle)=1$.
As the probabilities $P_i(|E_n\rangle)$ sum up to unity, the corresponding eigenstates are entirely localized in region (3), i.e., they are of type $|0,2_i\rangle$.
Above this energy, $P_2(|E_n\rangle)> 0$ indicates eigenstates $|1_i,1_j\rangle$ that (partially) populate region (2) where one boson is in each well.
To zeroth order\footnote{In this estimation of the lowest {\it total} energy of a $|1_i,1_j\rangle$ state, 
we neglected the tunneling coupling and the ground-state single-particle energy of the right well, which ---due to the large width--- is very small compared to  $\epsilon^{(sp)}_1$, cf. Eq.~(\ref{eq.sp_eig_energy}).}
(and in good agreement with the numerical results), this threshold energy is given by the single-particle ground-state energy in the left well $\epsilon^{(sp)}_1=3.79\times10^{-3}$ as given by (\ref{eq.sp_eig_energy}).
By the same argument, the first peaks in $P_1(|E_n\rangle)$ ---corresponding to eigenstates partially localized in region (1), i.e., with both bosons in the left well--- are observed only at about the energy of the two-particle ground state in the isolated left well $\epsilon_1^{}=0.0139$ [see the lowest curve in Fig.~\ref{fig.tp_box_spec_momenta}b)]. 

We cast this information into a density of states $\rho(E)$ of the environment, i.e., of the right well.
Later we will rather consider the {\it integrated} DOS 
\begin{equation}
\label{eq.IDOS}
n(E)=\int_{E_0}^E{\rm d}E \,  \rho(E)\,,
\end{equation}
which is less fluctuating.
Specifically, we are interested in the single-particle quasi-continuum ---where only one boson escaped from the left well--- and in the two-particle quasi-continuum ---where both particles are located in the broad right well.

The theoretical expressions for the single- $\rho^{sp}_{th}(E)$ and two-particle $\rho^{tp}_{th}(E)$ densities of states are derived in appendix \ref{app.DOS}, based on a non-interacting environment. 
This is, of course, a conceptually fundamental assumption to be tested in the following.
It seems plausible, however, to assume that the extremely short-ranged, repulsive interaction in Hamiltonian (\ref{eq.H_tp}) plays, if at all, a secondary role for the energy of the spatially extended continuum states.

In Fig.~\ref{fig.INT_DOS}b), we compare the theoretical predictions (\ref{eq.sp_N}), (\ref{eq.tp_N}) to the numerical results $n_{num}^{sp}(E)$ and $n_{num}^{tp}(E)$.
The latter are obtained from the integration of the corresponding curves in panel a) of the same figure, explicitly
 \begin{eqnarray}
 n_{num}^{sp}(E)&=&\sum_m\left[P_2(|E_m\rangle)\theta(E-E_m)\right]  \label{eq.sp_N_num} \,,\\
 n_{num}^{tp}(E)&=&\sum_m\left[P_3(|E_m\rangle)\theta(E-E_m)\right] \label{eq.tp_N_num} \,.
 \end{eqnarray}
For the single-particle quasi-continuum, the agreement between theoretical (\ref{eq.sp_N}) and numerical results (\ref{eq.sp_N_num}) is very good. The predicted energy gap and consecutive kinks are clearly observed, merely their position is slightly shifted to smaller energies with respect to the theoretical values given by Eq.~(\ref{eq.sp_eig_energy}). 
This shift results from the (tunnel-) coupling between the wells which is neglected in the derivation of
(\ref{eq.sp_N}).\footnote{\label{foot.Lamb_shift} The observed shift diminishes as we decrease the coupling between left well and quasi-continuum, e.g., by increasing the barrier height $V_0$.}

For the two-particle quasi-continuum, we find almost perfect agreement between $n_{num}^{tp}(E)$ and $n_{th}^{tp}(E)$, although the latter was derived for non-interacting particles.
For very large values of the interaction $U=1$ ---when the ground state is close to being fermionized---
we observe that the {\it integrated} DOS $n_{num}^{tp}$ is shifted to smaller values with respect to $n_{th}^{tp}(E)$ and thus a gap at $E=0$ opens.
This shift originates from the interactions in the right well [see Sec.~\ref{sec.tp_box} and Fig.~\ref{fig.tp_box_spec_momenta}b)] and vanishes for $r\rightarrow\infty$.
More importantly, the slope $\rho^{tp}(E)={\rm d}n^{tp}/{\rm d}E$, remains the same, i.e., the two-particle density of states in the quasi-continuum is unaffected by $U$. 
Furthermore, the gap decreases with increasing width $r$.
This consolidates our initial assumption that a contact interaction is of minor importance in the DOS of a very broad well which, ideally, stretches out to infinity.

\begin{figure*}[t]
\centerline{\includegraphics[width=1.95\columnwidth,keepaspectratio]{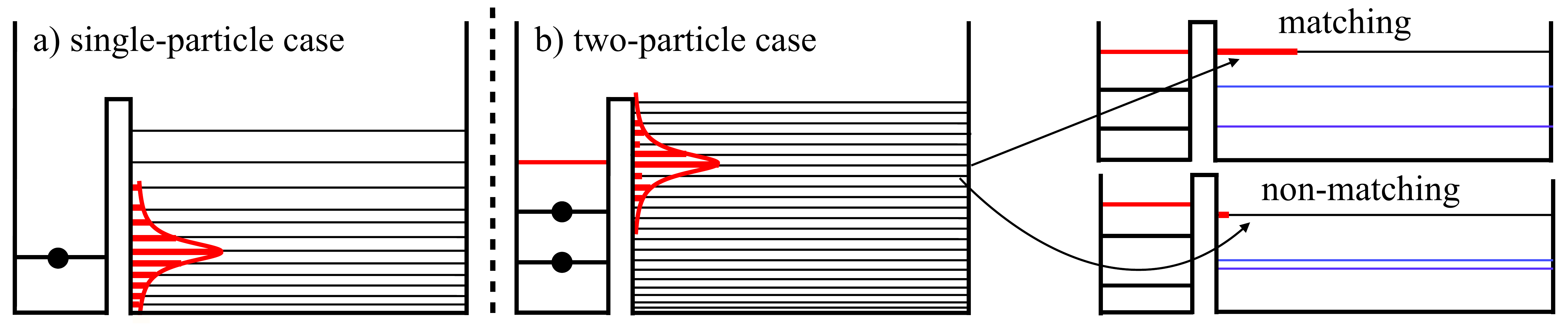}}
\caption{(Color online)
Sketch of tunneling decay in the single- and two-particle case.
{\it Left}: A single particle is prepared in the ground state of the left well (energy is marked by the bold black line).
For finite barrier height, the ground state couples to all eigenstates of the right well within the resonance width (red curve) and the magnitude of the coupling elements is indicated by the bold red lines.
{\it Middle}: Two interacting particles are prepared in the two-particle ground state (red) of the isolated left well at energy $\epsilon_1^{}=(k^{}_{1,1})^2+(k^{}_{2,1})^2$ (\ref{eq.E_tp}).
The energies $(k^{}_{i,1})^2$ corresponding to the momenta $k^{}_{i,1}$ (\ref{eq.k_1_2}) are marked in black.
For finite barrier height, and assuming mainly uncorrelated tunneling of the particles, the two-particle ground state {\it does not} couple to all two-particle eigenstates of the right well within the resonance.
Instead, the overlap (indicated by the bold red lines) is appreciable only for those states which are composed of single-particle eigenstates of the right well with single-particle momentum components which match those $k^{}_{i,1}$ of the initial state. 
{\it Right}: The top panel illustrates a \lq matching{\rq}  eigenstate  (black) of the right well, i.e., a large overlap, with corresponding (single-particle) energies of the right well marked in blue.
The bottom panel illustrates a \lq non-matching{\rq} eigenstate of the right well, where the corresponding single-particle energies 
(blue) are highly off-resonant with respect to the energies $(k^{}_{i,1})^2$ of the left well (black).  
}
\label{fig.LDOS_scheme}
\end{figure*}

According to our strategy which proved successful in the single-particle case of Sec.~\ref{ssec.3c}, the DOS together with the participation ratios PR and PR$_i$ should yield the dynamical decay rate $\gamma$ of, say, the particle number $N_{\ell}(t)$.
Inspection of the wave function's expansion coefficients $|c_n|^2$ in the two-particle eigenstates of the double well [see Eq.~(\ref{eq.c_n})] in Fig.~\ref{fig.INT_DOS}c) unravels a fundamental difference with respect to the single-particle case.
In contrast to the latter case (see the inset of Fig.~\ref{fig.sp_P_Left_PR}),
only a {\it fraction} of the $c_n$ take a finite value, within the relevant energy window around the two-particles ground-state energy $E=\epsilon_1=0.0139$.
As a result of the vanishing $|c_n|^2$ the two-particle DOS (\ref{eq.tp_N}) overestimates the number of {\it effectively} contributing states. 
This prevents proceeding as in the single-particle case through Eqs.~(\ref{eq.gamma_sp}),~(\ref{eq.rho_sp}): 
To identify  $\gamma$ with the quotient of the participation ratio and (in this case {\it overestimated}) DOS would {\it underestimate} the decay rate $\gamma$.
Nonetheless, we extract a resonance width 

\begin{equation}
\label{eq.gamma_tp_lorentz}
\gamma=5.9\times10^{-4}
\end{equation}
from the central 50\% \footnote{Specifically, we evaluated $\gamma$ from the condition $\int^{x_0+\gamma/2}_{x_0-\gamma/2} |c_n|^2 dE_n=0.5$.
Here, the center (i.e. the median) of the distribution, $x_0$, is determined by $\int^{x_0}_{-\infty} |c_n|^2 dE_n=0.5$.} of the expansion coefficient's distribution $|c_n|^2$, see Fig.~\ref{fig.INT_DOS}c), and comment on its role in the dynamical evolution in the next section.

What, at first sight, appears to be a further complication, {\em de facto} much elucidates the intricate tunneling mechanism: 
We have found that the vast majority of vanishing components within the resonance width originate from quasi-continuum states that {\it do have} the appropriate two-particle energy of the right well $\epsilon_n\approx \epsilon_i^{(sp)} +\epsilon_j^{(sp)}$ %
\footnote{If we ignore the influence of the repulsive interaction in the environment as above, the true two-particle energy $\epsilon_n$ can be replaced by a sum of two single-particle energies of the right well $\epsilon_j^{(sp)}$, given by Eq.~(\ref{eq.sp_eig_energy}), but with $\ell$ replaced by $r$.}
but their constitutent single-particle energies $\epsilon_j^{(sp)}$ do not match the momenta $k_{i,j}$ that enter the energy $\epsilon_1$ (\ref{eq.E_tp}) of the initial state $|\psi(0)\rangle$.
This demonstrates that ---unlike in the single-particle case--- it is not enough to consider the total energy $\epsilon_n$ of the initial (two-particle) state alone, as the greater part of the energetically matching two-particle states of the right well actually do not couple to the initial state, due to the mismatch between the underlying single momenta.
Put differently, tunneling decay in the two-particle case is not about resonances between the {\it total} two-particle energies but, rather
requires the {\it individual} matching of the two momenta $k_{i,1}$ in the left well to the momenta in the quasi-continuum.

In that sense, we are dealing with a case of single-particle tunneling, for which the inter-particle correlations play a minor role.
At the same time, this matching principle ---which we schematically illustrate in  Fig.~\ref{fig.LDOS_scheme}--- is of genuine two-body nature as it crucially relies on the individual properties (like, e.g., increased energy) of the two-particle wave vectors.
To properly interpret the illustration, we note that the assumption of uncorrelated single-particle tunneling implies that the initial $|2_1,0\rangle$-state is not directly converted into two-particle quasi-continuum states, but via intermediate states of type $|1_1,1_i\rangle$.

Let us summarize the main message of the spectral analysis:
For moderate values of the width $r$, isolated resonances exist in the PR and pair-wise two-particle tunneling from the initially prepared ground state $|\psi(0)\rangle$ can be expected for {\it appropriate choices} of the system parameters $r$, $\ell$, and $U$.
In the quasi-continuum case of $r\rightarrow\infty$ and repulsive interactions, resonances strongly overlap, compare Fig.~\ref{fig.PR_i}.
Although only a fraction of the two-particle continuum states was found to actually support the particle decay, the growing PR$_2$ and PR$_3$ indicate that ---irrespective of $r$--- the initial state couples to both, the single- and the two-particle continuum.\footnote{We remark that this is in perfect agreement with Fig.~\ref{fig.INT_DOS}b) which shows that, at the energy $\epsilon_1=0.0139$ of the initial two-particle ground state in the left well, $|0,2_i\rangle$- {\it and} $|1_1,1_i\rangle$-type states are available in the continuum, i.e, both are in resonance with $|\psi(0)\rangle$.}

We therefore develop the intuition that states which are transformed into each other by one single-particle process are considerably more strongly coupled than those which require a second-order process.
In total, we thus expect (second-order) pair-wise loss to be less generic than uncorrelated first-order tunneling to the quasi-continuum.

%=== INTERACTING PARTICLE LOSS DYNAMICS===
%
\section{Particle loss of two interacting bosons II: Dynamics}
\label{sec.tp_asdw_dyn}
\begin{figure*}[th]
\centering
\includegraphics[width=0.975\columnwidth,keepaspectratio]{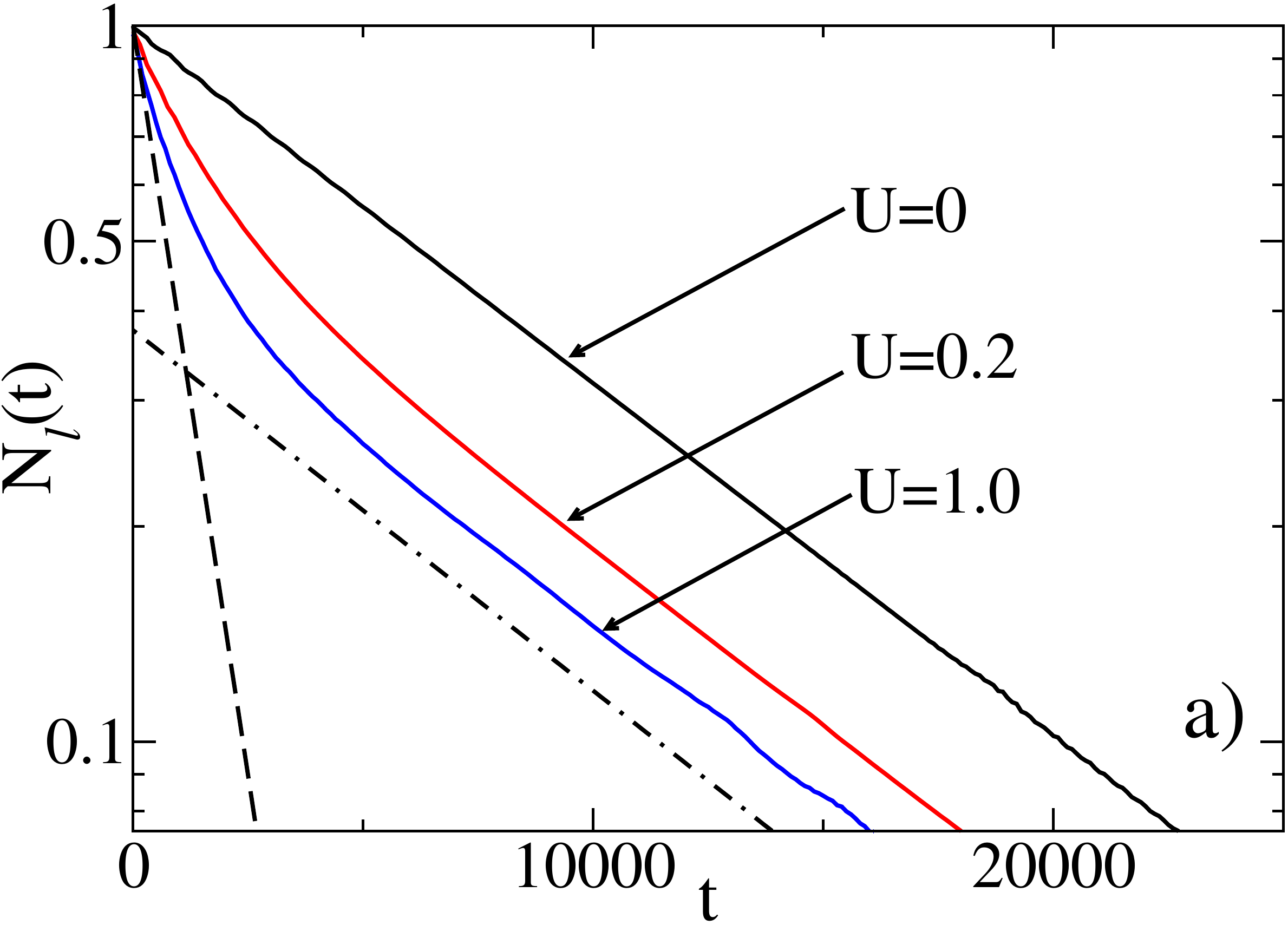}\hfill
\includegraphics[width=0.957\columnwidth,keepaspectratio]{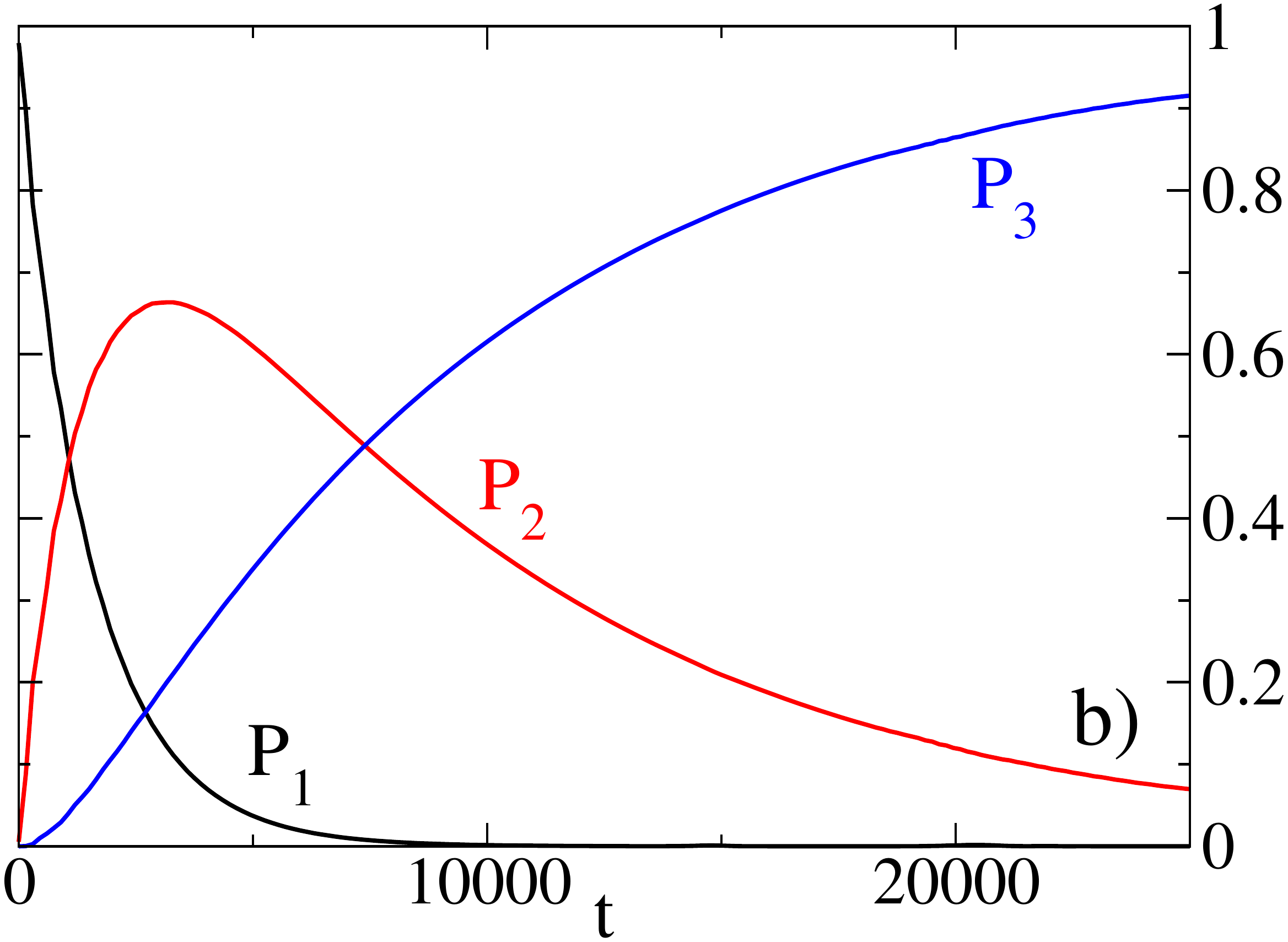}
\caption{Particle loss of two interacting bosons in an asymmetric double-well potential, in which the broad right well emulates the continuum.
The time evolution is obtained from diagonalization of (\ref{eq.H_tp}) with $l=51$, $b=2$, and $r=3000$, and the particles are initially prepared in the ground state
of the isolated left well (see Sec.~\ref{sec.tp_box})
{\it Left:} The normalized particle number $N_{\ell}(t)$ (\ref{eq.N_L}) in the left well, versus time $t$, for various values of the inter-particle interaction strength $U$.
For vanishing interaction, an exponential decay is observed.
As the interaction increases, pronounced deviations from the interaction-free exponential behavior arise. 
To guide the eye, the dash-dotted (dashed) line indicates an exponential decay with a rate corresponding to the single-particle ground (single-particle ground plus first excited) state [see Eqs.~(\ref{eq.gamma_sp1}),~(\ref{eq.gamma_sp2}), and (\ref{eq.P_1_limits})].
{\it Right:} For the exemplary interaction strength $U=0.2$, we plot the probabilities $P_i(t)=P_i(|\psi(t)\rangle)$ to find the particle in the three regions (1)-(3) of configuration space corresponding to two, one, and zero bosons populating the left well, respectively [see Eqs.~(\ref{eq.P_1})-(\ref{eq.P_3}) and Fig.~\ref{fig.setup}b)]: 
The strong initial increase of $P_2(t)$ compared to $P_3(t)$ indicates that the particles predominantly tunnel in the sequence (1)$\rightarrow$(2)$\rightarrow$(3).
That is, they tunnel {\it sequentially} and not as a correlated boson pair.
}
\label{fig.N_P_i_vs_time}
\end{figure*}

Up to this point we have not discussed the {\it time evolution} of two-particle loss, a topic which has only recently gained momentum.
First experiments range from atom loss off large BECs, induced by electron scattering \cite{WLGKO09} or photo ionization \cite{SBK10} to the tunneling-induced loss in highly controlled few-fermion systems \cite{ZSLWRBJ12}.
Only a handful of theoretical works treat the {\it full-fledged microscopic} decay problem considered here, and they rely on the propagation of the time-dependent Schr\"odinger equation either by direct integration \cite{KB11}, by combining a matrix-product state approach with Bose-Hubbard-like chains \cite{GC11},  or via multi-configurational Hartree-Fock methods \cite{LSAMC09,LSSAC12}.

In this section, we complement the results of our spectral analysis by the associated time evolution which we obtain by the spectral decomposition (\ref{eq.psi_t}) of the initial state (\ref{eq.psi_tp}) [but now for the two-particle quantities, resulting from the exact diagonalization of (\ref{eq.H_tp})], and comment on existing results as we proceed.
We use the same initial state as in the spectral analysis (see Sec.~\ref{sec.tp_phys_obs}), and the same width $r=3000$ of the right well, which we also considered in our analysis of the single-particle dynamics in Sec.~\ref{sec.spp}.

\subsection{Decay of the particle number $N_\ell(t)$ and the two-particle probabilities $P_i(t)$}
\label{sec.tp_dyn_N_of_t}
We start the discussion with an experimentally easily measurable quantity: the normalized number of bosons $N_\ell(t)$ in the left potential well  (\ref{eq.N_L}).
This quantity is plotted in Fig.~\ref{fig.N_P_i_vs_time}a), on a semi-logarithmic scale, for various values of the inter-particle interaction strength $U$.
For $U=0$, $N_\ell(t)$ decays exponentially with the decay rate of the single-particle ground state $\gamma_1^{(sp)}$ (\ref{eq.gamma_sp1})
(indicated by the dotted line).
This is not surprising since, in the interaction-free case, the system should effectively reduce to a single-particle problem.\footnote{Effects that arise from the indistinguishability of (non- and infinitely strongly interacting) particles are discussed, e.g., in \cite{DDGR06,CM11,Campo11,PSD12}.}
As $U$ increases, however, pronounced deviations from a (straight-line) exponential behavior arise, especially for times $t<5000$.
The second striking feature is that all curves asymptotically assume the same slope.

Insight is provided by the two-particle probabilities $P_{i=1,2,3}(t)=P_i(|\psi(t)\rangle)$ which ---by virtue of the definition (\ref{eq.N_L})--- determine $N_{\ell}(t)$. 
For the sake of visual clarity, we present the results for the exemplary value of $U=0.2$ in Fig.~\ref{fig.N_P_i_vs_time}b).
Initially, the probability $P_1(t)$ of finding both bosons in the left well drops.
At the same time, the probability $P_2(t)$ that one particle is located in the left and one in the right well increases.
Only after that does the probability of finding both bosons in the right well, $P_3(t)$, rise.
The key observation \cite{KB11} is that $P_2(t)$ becomes considerably large, independently of the interaction strength $U$.
We conclude that, to a large extent, the bosons do not leave the trap as a pair ---corresponding to a direct transition between regions $(1)\rightarrow(3)$ ---but sequentially leave the trap, i.e., through regions $(1)\rightarrow(2)\rightarrow(3)$.
This confirms our prediction based on the spectral analysis of the preceding section which we further substantiate in the following:
Namely, for $U\ge0$, the predominant loss mechanism is uncorrelated, i.e. independent, tunneling of the bosons \cite{LSAMC09,KB11,LSSAC12} {\it albeit} the presence of interactions.

With the help of the $P_i(t)$, we can now readily explain the short- and long-time behavior of the particle number $N_\ell(t)$.
The former will be essentially determined by the probability $P_1(t)$ which (in agreement with \cite{KB11}) is found to decay exponentially for {\it all} values of $U$.
We remind the reader that the initial state of the two bosons $\psi_{k_{1,1}k_{2,1}}(x,y)$ is characterized by the two wave vectors $k_{1,1} $ and $ k_{2,1}$.
Encouraged by the result of our spectral analysis and the behavior of the $P_i(t)$, we assume an independent particle picture as in \cite{KB11}, and associate a separate decay constant $\gamma_{k_{i,1}}$ with each wave vector.
That is, the probability $P_1(t)$ of finding both particles in the left well is given by the product 
\begin{align}
\label{eq.P_1_indep_pic}
P_1(t)=e^{-\gamma^{U}t}=e^{-\gamma_{k_{1,1}}t}\,e^{-\gamma_{k_{2,1}}t} \, ,
\end{align}
where the $k_{i,1}$ depend on $U$ via Eq.~(\ref{eq.k_1_2}).

Let us first consider the trivial case $U=0$.
Here, both particles are in the single-particle ground state $|\chi_1^\ell\rangle$, and the decay is determined by the corresponding single-particle rate given by (\ref{eq.gamma_sp1}), i.e., $\gamma_{k_{1,1}}=\gamma_{k_{2,1}}=\gamma_1^{(sp)}$.
For intermediate values $0<U<\infty$, the $U$-dependence of the rates can be inferred from Fig.~\ref{fig.tp_box_spec_momenta}a):
The momentum $k_{2,1}$ is largely independent of $U$ and given by the single-particle ground state momentum, hence we have
$\gamma_{k_{2,1}}\approx\gamma_1^{(sp)}$.
In contrast, the momentum $k_{1,1}$ ---and thus the associated energy $k_{1,1}^2$--- grows with $U$, leading to an increase of  $\gamma_{k_{1,1}}$ (since the effective barrier height seen by the particle is decreased).
This growth continues until, at large values of $U$, the system is fermionized and $\gamma_{k_{1,1}}$ takes the value of the decay rate $\gamma_2^{(sp)}$ associated with the single-particle first excited state (\ref{eq.gamma_sp2}).
In total, the decay of $P_1(t)$ therefore increases with $U$ and takes the limiting cases:
\begin{equation}
P_1(t)=\begin{cases}  e^{-2\gamma_1^{(sp)}t}\quad &U=0, \\
 e^{-[\gamma_1^{(sp)}+\gamma_2^{(sp)}]t}\quad &U=\infty \, 
\end{cases} \, .
\label{eq.P_1_limits}
\end{equation}
One thus finds that, for short times $t$, the (for non-vanishing $U$ larger) rate $\gamma_{k_{1,1}}$ dominates the time evolution of $N_{\ell}(t)$, and defines an upper limit ($\gamma_1^{(sp)}+\gamma_2^{(sp)}$) for the initial decay rate [indicated by the dashed line in Fig.~\ref{fig.tp_box_spec_momenta}a)].

Concerning the long-time evolution, we {\it assume} that, after the first boson left the trap, the remaining one populates the single-particle ground state.\footnote{This assumption is substantiated further on, in Sec. \ref{sec.tp_dyn_reduced_quant}, in the analysis of the reduced density matrix in energy space.}
Hence, this second boson decays with the single-particle decay rate $\gamma_1^{(sp)}$.
Put differently, the loss out of region (2) is governed by $\gamma_1^{(sp)}$, i.e., 
\begin{align}
\label{eq.P_2_indep_pic}
P_2(t\gg1)\propto e^{-\gamma_1^{(sp)}t}.
\end{align}
In agreement with our numerical simulations, the $U$-independent, smaller decay rate $\gamma_1^{(sp)}\approx\gamma_{k_{2,1}}$ (dash-dotted line) governs the asymptotic behavior of $N_l(t\gg1)$ for all values of $U\ge0$.

From the preceding discussion, we cannot but conclude that the assumption of uncorrelated tunneling very well reproduces the observed decay of the particle number inside the left well.

We point out that our results on $N_\ell(t)$ and $P_1(t)$ excellently agree with those of Ref.~\cite{KB11}, in which a $\delta$-like potential barrier was considered.
In contrast, Refs.~\cite{LSAMC09,Lode09_corig,LSSAC12} do not mention such non-exponential decay of the particle number $N_\ell(t)$.
Neglecting the different trap geometry, the parameters used in the main part of \cite{LSSAC12}  roughly correspond to interatomic interactions of the order of $U\approx0.05$;
inspection of Fig.~\ref{fig.tp_box_spec_momenta}a) reveals that, for $U\approx0.05$, the difference in the wave vectors $k_{i,1}$ is comparatively small.
Thus also the rates $\gamma_{k_{1,1}}$ and $\gamma_{k_{2,1}}$ barely differ, making it difficult to observe non-exponential decay.
In the associated  {\it Supporting Information} of \cite{LSSAC12}, the time evolution for two bosons and a seven-fold larger interaction strength is reported in Fig.~S2.
A careful re-analysis of the data (semi-logarithmic plot) reveals that the decay indeed consists of two exponentials, as discussed above, leaving no inconsistencies.

We note in passing that we determined the decay rate of $P_1(t)$ by an exponential fit to be
\begin{equation}
\gamma^{U=0.2}=6.8 \times 10^{-4}\,,
\label{eq.gamma_U=0,2_time_evo}
\end{equation}
which is in good agreement with the spectrally extracted width (\ref{eq.gamma_tp_lorentz}) of the Lorentzian curve in Fig.~\ref{fig.INT_DOS}c).
As we explain in Appendix~\ref{app.P_s_w_c},
this indicates that for our setup, the probability $P_1(t)$ corresponds to the survival probability $P_{\rm surv}=|\langle\psi(0)|\psi(t)\rangle|^2$.

\begin{figure*}[t]
\centering
\includegraphics[width=0.974\columnwidth,keepaspectratio]{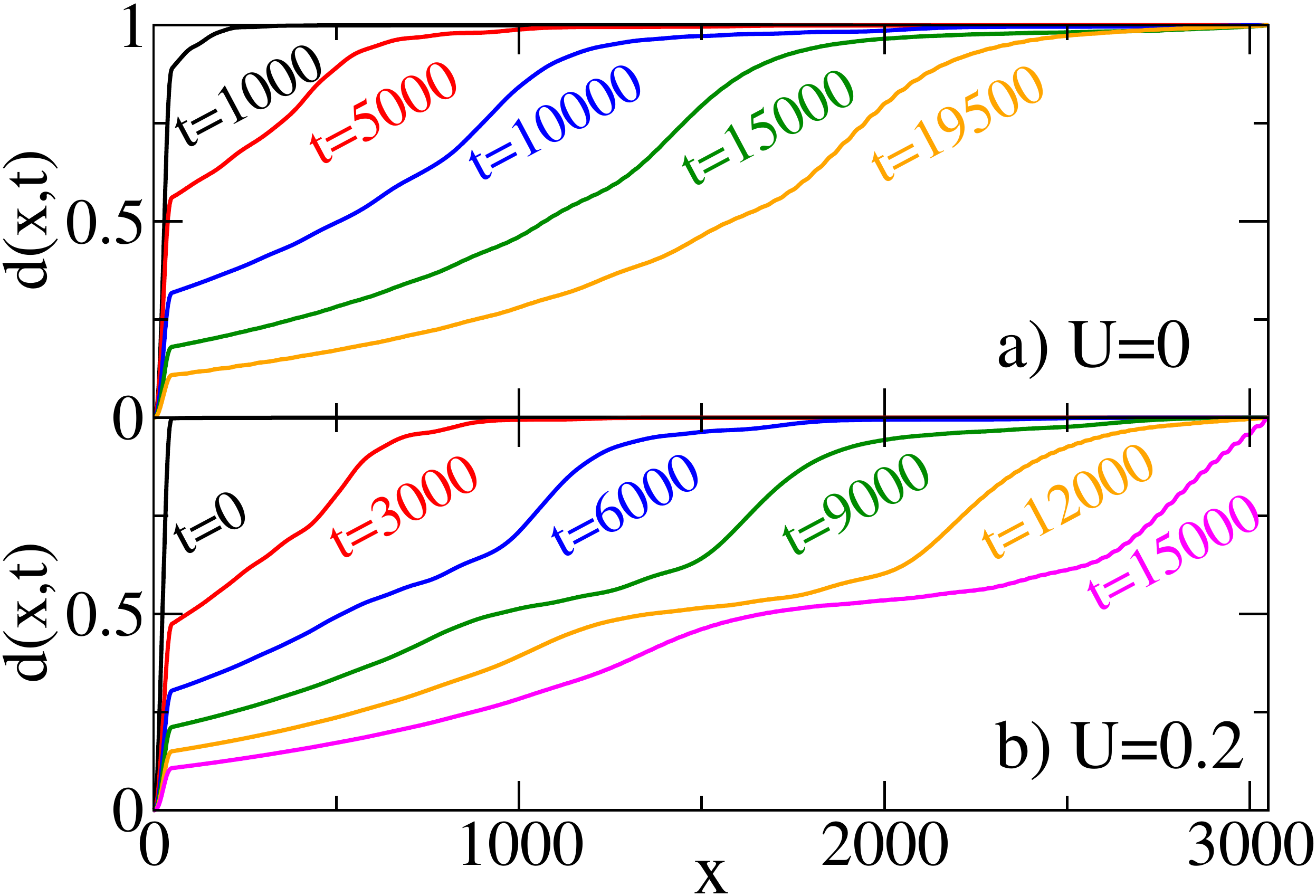}\hfill
\includegraphics[width=0.97\columnwidth,keepaspectratio]{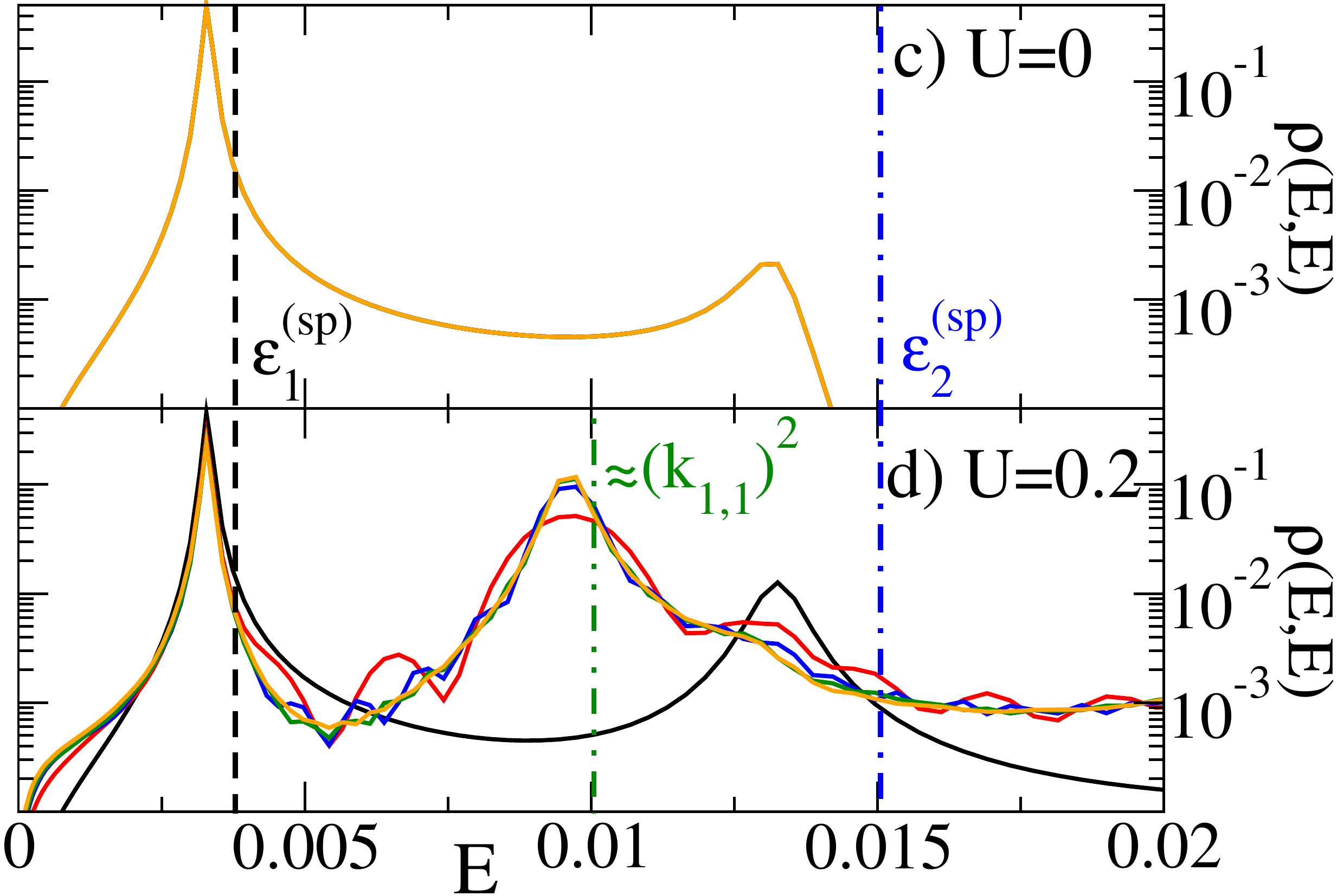}
\caption{(Color online) Reduced single-particle quantities of the two boson decay.
{\it Left:} Integrated diagonal $d(x,t)$ (\ref{eq.int_rho_red}) of the reduced single-particle density matrix $\rho_{\rm red}(x,x,t)$  versus position $x$, for different evolution times $t$ and inter-particle interaction strengths $U$, for otherwise same parameters as in Fig.~\ref{fig.N_P_i_vs_time}. 
For vanishing interactions, the particles leave the left well independently and with the same probability.
Hence, one observes a continuous increase of $d(x,t)$.
In contrast, for $U=0.2$, the bosons tunnel to the right well at different rates and different asymptotic momenta.
This leads to a spatial separation in the quasi-continuum, manifest in the plateau around the interval $x\in[1500,2400]$ (for $t=15000$).
{\it Right:} The diagonal $\rho_{\rm red}(E,E,t)$ (\ref{eq.rho_red_energy}) of the reduced single-particle density matrix in energy representation versus the energy $E$, for the same parameters and color code as in panels a) and b).
In the interaction-free case $U=0$, one observes a dominant Lorentzian-shaped peak at $\epsilon^{(sp)}_1$ (dashed vertical line) which comprises about 98\% of the total probability and a smaller peak at $\epsilon^{(sp)}_2$ (dash-dotted vertical line).
The reduced probability density $\rho_{\rm red}(E,E,t)$ does not change in time, hence all curves fall on top of each other. 
For repulsive interactions, the {\it initial} distribution is qualitatively similar to the case $U=0$.
During intermediate times $t<6000$, the interaction energy is transformed into kinetic energy and the distribution $\rho_{\rm red}(E,E,t)$ changes.
As a result of this conversion process, two peaks emerge, located at the
energy associated with the two wave vectors $k^2_{1,1}\approx\epsilon^{(sp)}_1$ and $k^{2}_{2,1}\approx\epsilon^{}_1-\epsilon^{(sp)}_1$ (dash-double-dotted vertical line) which constitute the initial state.
Each of the peaks in the asymptotic distribution comprises 50\% of the probability, which corroborates the picture of two asymptotically independent particles.
}
\label{fig.tp_diag_rho_red_vs_time}
\end{figure*}

\subsection{Reduced density matrix in coordinate and energy space}
\label{sec.tp_dyn_reduced_quant}
We now wrap up the discussion of the decay dynamics with two complementing reduced single-particle observables.
These are potentially more easily accessible in the experiment than, e.g., the two-particle probabilities $P_i(t)$.
On the other hand, they add to the understanding of the time-resolved tunneling process.
We start with the integrated diagonal $d(x,t)$ of the reduced density matrix $\rho_{\rm red}(x,x)$, i.e.,
\begin{equation}
d(x,t) = \int_0^x {\rm d}x' \ \rho_{\rm red}(x',x',t)\, ,
\label{eq.int_rho_red}
\end{equation}
which denotes the probability of finding ---at a given time $t$--- a boson in the interval $[0,x]$ of configuration space.
Consequently, as long as $d(x,t)$ increases with $x$ (for fixed $t$), there is a non-vanishing probability that a boson resides at the position $x$.
As will become clear in the next paragraph, when seen as a function of time, $d(x,t)$ visualizes how the bosons leave the left well.
We finally note that $d(\ell,t)=N_\ell(t)$, as the left well stretches from $x=0$ to $x=\ell$.

In the left panels of Fig.~\ref{fig.tp_diag_rho_red_vs_time}, we plot $d(x,t)$ versus the position $x$, for various times $t$ and two representative values of $U$.
To familiarize ourselves with the quantity at hand, we first consider the interaction-free case $U=0$, shown in panel b).
For example, $d(51,15000)\approx0.2$, that is, at time $t=15000$, about $80\%$ of the particles have tunneled through the potential barrier located at $x=51$.
More importantly, $d(x>51,t)$ shows a monotonic increase in $x$ for all (fixed) times.
Since for $U=0$, the two bosons are in the same single-particle (ground) state [see Eq.~(\ref{eq.psi_u=0})] and  do not interact with each other, this constitutes our reference case of two bosons leaving the left well with the same probability.

Contrast this with repulsively interacting bosons [$U=0.2$, panel b)]:
As time evolves, a plateau at $d(x,t)\approx1/2$ emerges, see, for example, the magenta curve ($t=15000$) around $x\in[1500,2400]$.
According to the definition of $d(x,t)$, this implies that the probability of finding a particle within this plateau is about zero.
Put differently, with probability one half one finds a boson between $x=0$ and $x=1500$ and equally likely a boson can be detected between $x=2400$ and $x=3000$.
This behavior nicely fits our picture developed above, in which the bosons {\it independently} leave the trap but (for $U>0$) with {\it different} loss rates $\gamma_{k_{i,1}}$.
For even larger times, this plateau is stretched and the two particles are further separated, indicating that they travel at different speeds.

To corroborate this statement and to complement the discussion on the particles' spatial distribution, we finally discuss the fingerprint of the particle loss in energy domain ---let aside the advantage that the energy (i.e., the momentum) of a single particle may be easier to measure than its position.
For this purpose, we consider the reduced single-particle density matrix in energy representation 
\begin{equation}
\rho_{\rm red}(E,E',t) = T(E,x) \rho_{\rm red}(x,y,t) T^\dagger(E',y) \, ,
\label{eq.rho_red_energy}
\end{equation}
$T$ being the unitary basis transformation from configuration space to the energy domain.

In what follows, the reader is warned that the reduced quantities are associated with measurements on a single particle.
While even a single boson carries some information on the total system (due to symmetrization), the reduced quantities do not reflect correlations between the particles, e.g., the probability that both particles have the same energy, cannot be deduced from $\rho_{\rm red}$. 
Here, we focus on the diagonal elements $\rho_{\rm red}(E,E,t)$, i.e., the probability density of finding {\it one} particle with energy $E$.
In the right panels of Fig.~\ref{fig.tp_diag_rho_red_vs_time} we present $\rho_{\rm red}(E,E,t)$ versus $E$, for the same parameters as in panels a) and b) of the same figure.

Paralleling the discussion for $d(x,t)$, we take the interaction-free case [panel c)] as our starting point. 
Initially, the two bosons populate the left well and the two-particle initial state $|\psi(0)\rangle$ is characterized by the two momenta $k_{1,1}$ and $k_{2,1}$.
For $U=0$, the latter are identical and furthermore coincide with the momentum $k_1^{(sp)}$ of the single-particle ground state, see Fig.~\ref{fig.tp_box_spec_momenta}a).
Intuitively, one would thus expect to observe a peak in $\rho_{\rm red}(E,E,0)$ at the corresponding energy $\epsilon^{(sp)}_1$ (dashed vertical line).
Indeed, we identify one dominating, Lorentzian-shaped peak (comprising about $98\%$ of the probability\footnote{The probability comprised  by each peak is calculated via integration of $\rho_{\rm red}(E,E)$. As integration boundaries, we take the middle between the centers of neighboring peaks.}), located slightly below $\epsilon^{(sp)}_1$.
As we already remarked in Sec.~\ref{sec.spp}, this energy shift results from the coupling of the left well to the quasi-continuum and increases with the energy of the initial state [see also Footnote~\ref{foot.Lamb_shift}].
The remaining probability is contained in the second, inferior peak, situated close to $\epsilon^{(sp)}_2$ (dash-dotted vertical line).
We remark that this peak would as well appear in a truly single-particle calculation and results from a finite overlap of the initial state $|\psi(0)\rangle$ and quasi-continuum states that also involve excited single-particle states of the left well.
These minor corrections aside, intuition does not fail us.
More importantly than the position of the peaks, $\rho_{\rm red}(E,E,t)$ does not change its shape as time evolves:
All curves exactly fall on top of one another.
Accordingly, both bosons follow the same energy distribution at all times, that is, they move with the single-particle ground-state momentum through the quasi-continuum, as expected for the non-interacting case. 

How do repulsive interactions alter this picture [$U=0.2$, panel d)]?
In contrast to the case $U=0$, the initial momenta $k_{1,1}$ and $k_{2,1}$ no longer coincide with the single-particle momenta $k_i^{(sp)}$ (i.e., with the $\epsilon_i^{(sp)}$) and one is tempted to expect a direct manifestation thereof in the initial distribution $\rho_{\rm red}(E,E,0)$.
Yet, the difference in $\rho_{\rm red}(E,E,0)$ compared to the case $U=0$ seems marginal:
For $t=0$, we still observe two peaks at the same positions as before, save that the main peak centered around $\epsilon^{(sp)}_1$ now  comprises only about $90\%$ of the probability.\footnote{The largest part of the remaining probability is contained in the second peak [which is increased with respect to panel c)] while higher than first excited single-particle eigenstates contribute with considerably less weight.}
The reason why the $k_{i,j}$ do not emerge is that, at $t=0$, only single-particle eigenstates of the {\it left} well can be populated on the reduced single-particle level.
Only {\it after} the particles tunnel to the right well are quasi-continuously distributed states available, and every energy component can contribute to the (reduced single-particle) dynamics.
In accordance with the warning issued after Eq.~(\ref{eq.rho_red_energy}), $\rho_{\rm red}(E,E,0)$ rather reflects the interaction-induced finite contribution of $|\chi_2^\ell\rangle$ to the initial state $|\psi(0)\rangle$  than the (truly two-particle) momenta $k_{i,1}$.

The influence of the interaction becomes fully apparent as time evolves: 
The peak at $\epsilon^{(sp)}_1$ declines\footnote{This decline is less pronounced on the logarithmic scale, but can be inferred from the increasing second peak.} while the one at $\epsilon^{(sp)}_2$ entirely vanishes, and a new one centered around the energy $E=9.7\times10^{-3}$ emerges.
The change in the shape of $\rho_{\rm red}(E,E,t)$, clearly indicates that energy is being redistributed among the particles.
In the independent decay picture, the particle that remains in the left well populates the single-particle ground state, i.e., it has energy $\epsilon^{(sp)}_1 \approx (k_{2,1})^2$.
Due to energy conservation, the boson that tunneled out of the left well should carry the energy $\epsilon_1-\epsilon^{(sp)}_1\approx(k_{1,1})^2$.
Indeed, we can associate the emerging peak with the latter energy, while the first peak still corresponds to the ground state energy $\epsilon^{(sp)}_1$, in good agreement with the result of \cite{LSSAC12}.
Our interpretation that the two peaks represent the two bosons is corroborated by the fact that (for times $t>6000$) each peak comprises half of the probability.
We furthermore found that the energy widths of the two (asymptotically) Lorentzian peaks in $\rho_{\rm red}(E,E,t)$ approximately equal the two decay constants $\gamma^U$ and $\gamma_1^{(sp)}$ associated with the decay of the two particles.
All three observations concerning the position, integrated area, and width of the peaks in $\rho_{\rm red}(E,E,t)$ further substantiate the picture of uncorrelated, independent decay of the bosons.
Hence, $\rho_{\rm red}(E,E,t)$ represents an experimentally easily measurable quantity which bears substantial information on the tunneling decay.

Let us now take a closer look at the energy conversion which takes place during the tunneling process.
Initially, both bosons populate the left well, and interact with each other.
Once they tunnel to the right well, their  interaction energy is converted into kinetic energy.
This process lasts until $t\approx6000$, until when the second peak grows and takes a Gaussian shape [see the red curve in Fig.~\ref{fig.tp_diag_rho_red_vs_time}d)].
For larger times ($t\ge6000$), one may imagine that the boson with the larger decay rate has completely tunneled to the right well and is moving at a larger momentum $\approx k_{1,1}^{}$.
This is also approximately the time when the plateau in $d(x,t)$ appears, see panel b).
Then the bosons cease to interact, and the peak takes its asymptotic Lorentzian shape.
As a result, the interaction part of the total energy (\ref{eq.H_tp}) effectively vanishes at these large times and we find
\begin{eqnarray}
 \langle\psi(t)| H_{sp} \otimes {\hat 1} + {\hat 1} \otimes H_{sp}|\psi(t) \rangle = 2\ {\rm tr}(H_{sp}\,\rho_{\rm red}) = \epsilon_1^{} \nonumber \,.\\
 &&
\end{eqnarray}
That is, the total energy $\epsilon_1$, determined by the initial state, is given by twice the average single-particle energy which is certainly not the case for smaller times, when the particles still interact.

As a last remark, we have numerically confirmed that, for $r\rightarrow \infty$, the repulsive interactions between the bosons in the right well have almost no effect on the loss dynamics.\footnote{We set $U=0$ in the right well, and found almost no difference with respect to the curves of Fig.\ref{fig.N_P_i_vs_time}.}
Hence, it is well justified to ignore them, which is in perfect agreement with our results from Sec.~\ref{subsubsection.DOS_U>0}, where both, the single-particle and the two-particle densities of states of the quasi-continuum proved almost independent of $U$.
While we stress that this result was derived only for $\delta$-like interactions, the possibility of neglecting interactions in the environment greatly simplifies perturbative approaches to the loss dynamics, such as a master equation ansatz.

Summarizing this section, we have found that repulsive interactions strongly modify the loss dynamics and lead to a multi-exponential decay of the boson number. 
Notwithstanding, the dominant decay mechanism is found to be uncorrelated tunneling of single particles.
That is, in none of the studied examples was significant correlated two-particle loss observed.
Recalling the spectral analysis of the previous section, both single and two-particle tunneling were found to be energetically possible.
Yet the simple physical picture developed in Sec.~\ref{sec.tp_asdw_spec} turns out to be correct:  single-particle tunneling ---as a first-order process--- wins over the slower two-particle process of second order and thus constitutes the dominant tunneling mechanism.

\section{Summary and Discussion}
\label{sec.concl}

In the present work we studied the tunneling of two ultracold bosons, initially prepared in the left site of a one-dimensional double-well potential, as the width of the right well was gradually increased in subsequent realizations of the trap geometry.
Seen from the left well, a broad right well mimics an unconfined configuration space (i.e., an environment) with a dense quasi-continuum of states to which the particles escape.
Through the numerically exact diagonalization of the full two-body Hamiltonian, we could identify those quasi-continuum states which actually support this decay and expose the role played by the repulsive interactions.
We could also conclude that the many-body decay process is governed by independent tunneling of the bosons rather than by tunneling of a boson pair.
We briefly review the main points.

As a starting point and frame of reference, we first investigated the single boson case.
In the participation ratio PR of the initial state, we found a simple spectral tool, to predict particle dynamics: 
Its maxima with respect to the width $r$ of the right well, termed {\it resonances}, reliably indicate those trap geometries for which substantial tunneling occurs.
Still isolated at comparable widths of the two wells, the resonances increasingly overlap as the quasi-continuum $r\rightarrow\infty$ is approached.
This was identified as a spectral signature of the transition from tunneling oscillations to tunneling decay.
The rate of the corresponding exponential particle loss was extracted from the asymptotic growth of the PR and found to be in good agreement with the associated time evolution as well as with analytical results.

In the main part of this work, we then analyzed the case of two bosons, initially prepared in the two-body ground state of the isolated left well.
As a first major difference compared to the single-particle case, the participation ratio PR exhibits resonances of two {\it different} widths.
We introduced the concept of configuration-space sensitive participation ratios PR$_i$ and unequivocally identified narrow resonances with second-order pair-wise tunneling, and broad resonances with uncorrelated first-order tunneling.

In the quasi-continuum limit of a broad right well, the resonances of each kind were found to overlap, which implies that {\it a priori} both single-particle {\it and} two-particle states of the continuum are available for the loss dynamics.
This statement was corroborated by the excellent agreement of our analytically and numerically evaluated single- and two-particle density of states which furthermore proved to be independent of the inter-particle interaction $U$.

As a second major difference with respect to the single-particle decay, we showed that not all {\it energetically} allowed quasi-continuum states contribute to the tunneling, but only a small fraction thereof.
It was found that this fraction is determined by the condition that the single-particle energies, which underlie the quasi-continuum, match the energies associated with the two-particle momenta $k_{i,1}$ of the initial state.

We complemented our spectral analysis with the study of the associated time evolution.
In agreement with previous works \cite{KB11}, we found that depending on $U$, the initial decay of the boson number in the left well $N_\ell(t)$ fundamentally deviates from the   (interaction-free) exponential behavior, while the asymptotic decay of two bosons was found to be exponential and interaction independent.
The results were explained in terms of the interaction-dependent two-particle wave vectors $k_{i,1}$.
We furthermore showed that the essential characteristics of the two-particle loss process can be extracted from measurements of reduced single-particle quantities and elaborated on the conversion from potential to kinetic energy during the tunneling process.

In combining the results of the static (spectral) and dynamic analysis, we developed a simple and clear picture of the tunneling decay:
The repulsive interactions first and foremost modify the energy of the bosons which eventually determines their tunneling rate and final momenta.
That is, although the initial state is of truly two-body nature, the tunneling decay is well reproduced by associating the two-particle wave vectors $k_{i,1}$ with independent particles.
Furthermore, through appropriate choice of the system parameters, (second-order) tunneling of a boson pair may be enforced for {\it moderate} asymmetries of the trap where the resonances are still isolated.
The particle loss, however, is dominated by single-particle tunneling which (as a first-order process) wins over the slower two-particle process of pair-tunneling.

Experimentally, quantum systems of few ultracold particles have recently been realized for arbitrary values of the inter-particle interactions \cite{SZLOWS11} and the loss process of the {\it first} particle has already been monitored \cite{ZSLWRBJ12}.
Although even box shaped potentials seem experimentally feasible \cite{EWARWD10,HRMB09,MSHCR05}, our observations should not qualitatively depend on the specific form of the trapping geometry. 
Hence, an experimental investigation of the few-body decay is realizable with state-of-the-art technologies.
Besides the decay, the isolated resonances in the participation ratio for comparable widths of the two wells demonstrate an experimentally feasible way to address ---in a controlled way--- two- and single-particle tunneling by changing the geometry of the trap, rather than tuning additional magnetic fields.

In closing, let us outline future directions of this research area. 
First, the fraction of pair-tunneling in the loss dynamics ---as low as it may be--- should be determined in a future study together with a prescription to experimentally detect it.
Second, due to their inherent symmetry, bosonic systems exhibit particle bunching \cite{HOM87}, even for massive, non-interacting bosons \cite{JMHVKSPCBAW07,TTMMB10}.
One may thus very well ask to what extent the interaction-induced fermionization process observed in the present setup modifies the bunching behavior.
Intimately related is the dependence of the dynamics on initially prepared states different from the ground state of the left well, e.g. excited states or non-eigenstates \cite{GC11,lewin10}.
Third, we here considered a $\delta$-like interaction which represents the simplest two-body interaction.
What will happen when more complex (e.g., long-range, Coulomb) interactions come into play?
First results exist for the so-called escape dynamics \cite{TS11} (a precursor of tunneling decay in which the confining potential is suddenly switched off and particles escape to free space).
A recent study \cite{MSS12} indicates that charged ultracold bosons in continuous potentials may exhibit quantum chaotic behavior (much like that observed in quantum lattice models, see, e.g. \cite{BK03}), which may alter as well the decay process \cite{FKCD01}.
Similarly, the study of attractively bound boson-pairs \cite{Hao:2006dp} would be of both experimental and theoretical interest.
For example, will the tunneling break the bonding or will the particles tunnel as a pair?

We finally point out a long-term objective of eminent computational relevance.
While the decay of a small number of bosons can still be treated exactly \cite{LSSAC12} or within a Bose-Hubbard-type modeling \cite{GC11}, the microscopic dynamics of larger boson numbers is beyond numerical reach.
The observed (predominantly) uncorrelated single-particle tunneling decay might advance the development of perturbative approaches, such as a master equation.
Without resorting to heuristically introduced approximations, one should base the master equation ansatz on the energy-dependent loss rates described above and compare the outcome to existing results.

\begin{acknowledgments}
We thank Iva B\v{r}ezinov\'a, Joachim Burgd{\"o}rfer, and Dimitry Krimer for helpful discussions.
We acknowledge support by DFG {\it Research Unit 760} and the EU COST Action MP1006 'Fundamental Problems in Quantum Physics'.
S.H. is grateful for support by the Konrad-Adenauer-Stiftung.
\end{acknowledgments}

\bibliography{../../../ASDW}

%%%%%%%%%%%%%%%%%%%%%%%%%%%%%%%%%%%%%%%%%%%%%%%%%%%%%%%%%%%%%%%%%%%%%%%%%%%%%%
\appendix

%=== APPENDIX - TWO-BODY HAMILTONIAN ===
%
\section{Two-body Hamiltonian}
\label{sec.app}
The most general two-boson Hamiltonian reads:
\begin{align}
H =& \int dx \ \Psi^{\dagger}(x) H_{sp} \Psi^{}(x) \nonumber \\
    &+ \int dx \int dy\ \Psi^{\dagger}(x)\Psi^{\dagger}(y) W(x,y) \Psi^{}(x) \Psi^{}(y) \, ,
\label{eq.H_mp}
\end{align}
where $\Psi^{(\dagger)}(x)$ are bosonic field operators which annihilate (create) a boson at position $x$ and $H_{sp}$ is the single-particle Hamiltonian defined in Eq.(\ref{eq.H_sp}).
The second term accounts for the interaction $W(x,y)$ between the two particles with coordinates $x$ and $y$, respectively.
We construct the two-particle Hamiltonian by expanding the bosonic field operators in (\ref{eq.H_mp}) in terms of the eigenfunctions $E^{(sp)}_j(x)=\langle x|E^{(sp)}_j\rangle$ of the single-particle Hamiltonian $H_{sp}$, i.e.,
\begin{equation}
\Psi^{}(x) = \sum_j a_j E^{(sp)}_j(x) \, .
\label{eq.expansion}
\end{equation}
This yields
\begin{equation}
H_{tp} = \sum_j \ E^{(sp)}_j \ a^{\dagger}_ja_j + \sum_{i, j, l, m} g_{ijlm} a^{\dagger}_i a^{\dagger}_j a_l a_m \, ,
\label{eq.H_tp_2}
\end{equation}
where $E^{(sp)}_j$ are the single-particle eigenenergies and $g_{ijlm}$ is the matrix element that accounts for the contact interaction 
$W(x,y) = U\delta(x-y)$ (see discussion after (\ref{eq._V_sp}) in the main text) between the particles, i.e., 
\begin{equation}
g_{ijlm} = U \int dx \  \left(E^{(sp)}_i(x)E^{(sp)}_j(x)\right)^{*} E^{(sp)}_l(x) E^{(sp)}_m(x) \,  .
\label{eq.g_ijkl}
\end{equation}
In our numerical treatment, we retain the lowest $n_{cut}$ single-particle eigenfunctions leading to a two-particle Hilbert space of dimension
$(n_{cut})(n_{cut}+1)/2$ which is spanned by the basis $\{{\cal S}( |E^{(sp)}_m\rangle \otimes |E^{(sp)}_n\rangle) \}$.  Here $\{ |E^{(sp)}_m\rangle\}$ is the
single-particle energy eigenbasis, ${\cal S}$ is the symmetrization operator, and $n\le m =1,...,n_{cut}$.
In this approach, convergence is reached when the results do not change upon further increase of $n_{cut}$.
Typically, several hundred single-particle states are taken into account which enables us to study a broad range of repulsive interactions $U$. 
We note that this treatment goes far beyond the single-band approximation often assumed in the context of ultracold atoms, e.g. within the Bose-Hubbard approach \cite{JBCGZ98}.

%=== APPENDIX - ANALYTICAL SINGLE-PARTICLE ===
%
\section{Analytical treatment of single-particle loss}
\label{app.ana_sp}
For an infinitely broad right well ($r\rightarrow\infty$), we can readily formulate the single-particle loss as a scattering problem.
In each of the three intervals I, II, and III of configuration space [see Fig.~\ref{fig.setup}a)], we make a (textbook-like) plane-wave ansatz for the single-particle wave function:
\begin{equation}
\psi(x)=\begin{cases}
A \exp(-ikx) + B \exp(i k x) & x \in  {\rm I} \\
C \exp(- \kappa x) + D \exp( \kappa x)  &x  \in {\rm II}\quad ,\\
F \exp(-ikx) + G \exp(i k x) &x  \in {\rm III}
\end{cases}
\end{equation}
with $\kappa^2 = V_0 - k^2$.
The postulated continuity of $\psi(x)$ and its first derivative at the borders of the intervals yields four equations for the six coefficients $A$ to $F$.
We are seeking purely outgoing solutions and thus fix the amplitude $F(k)=0$.
The latter equation has solutions $\tilde{k}^{(sp)}_{n}$ that lie in the complex plane, leading to complex energies $\tilde{E}^{(sp)}_n=(\tilde{k}^{(sp)}_{n})^2$ whose imaginary part corresponds to the decay rate of the $n$-th eigenstate $|\chi_n^{\ell}\rangle$ of the isolated left well.

%%=== APPENDIX - ANALYTICAL TWO-PARTICLE ===
\section{Two- and single-particle density of states}
\label{app.DOS}

In this appendix, we derive theoretical expressions for the single- and
two-particle density of states in the quasi-continuum, which correspond
to one and two particles having escaped from the left well, respectively.
The comparison to our numerical data is found in
Sec.~\ref{subsubsection.DOS_U>0}, see the discussion of Fig.~\ref{fig.INT_DOS}.

We assume that the single-particle quasi-continuum is formed by product
states $|\chi_n^{\ell}\rangle\otimes|\chi_m^r\rangle$ of the
single-particle eigenstates of the left and right well.
This assumption neglects the tunneling coupling and is thus strictly
valid only for uncoupled wells.
The interaction between bosons in the left and right well, on the other
hand, can be safely ignored since we are dealing with a contact
potential in Hamiltonian (\ref{eq.H_tp}).
According to Eq.~(\ref{eq.rho_sp}), the $|\chi_m^r\rangle$ determine the
integrated single-particle DOS of the right well.
Since we would like to obtain the integrated DOS as a function of the
{\it two-particle} energy $E$, we have to additionally take into account
the energy of the states $|\chi_n^{\ell}\rangle$.
By summing over all combinations $\{n,m\}$ with total energy below a
fixed value $E$, we obtain the integrated DOS of the single-particle
quasi-continuum,
\begin{equation}
n_{th}^{sp}(E)= \frac{r}{\pi} \sum_m\left[\sqrt{E-\epsilon^{(sp)}_m}\
\theta(E-\epsilon^{(sp)}_m)\right] \, ,
\label{eq.sp_N}
\end{equation}
where the $\epsilon^{(sp)}_m$ are the eigenenergies of the isolated left
well [see prior to Eq.~(\ref{eq.c_n})] and $\theta(\cdot)$ is the
Heaviside step function.
We note that the latter implies an energy gap of size
$\epsilon^{(sp)}_1$ and potentially also discontinuities (kinks) at
consecutive energies $\epsilon^{(sp)}_n$.

For the two-particle continuum, an analytical expression is more
involved, as the two-particle eigenenergies result from (\ref{eq.E_tp})
and the coupled equations (\ref{eq.k_1_2}).
Instead, let us assume for the nonce that the contact interaction plays
a secondary role as far as the energy is concerned, since the right well
consists of the spatially extended quasi-continuum states.
We shall discuss the validity of this assumption in the main text.
Then, for two {\it non-interacting} bosons, the associated two-particle
density of states is given by \cite{MFMR06,SW03}:
\begin{equation}
\rho_{th}^{tp}(E) = \frac{1}{2} \left( [\rho^{(sp)}\star\rho^{(sp)}](E) +  \frac{1}{2} \rho^{(sp)}\left( \frac{E}{2}\right) \right) \, ,
\label{eq.tp_DoS}
\end{equation}
where $\star$ indicates the convolution and $\rho^{(sp)}$ is the single-particle density of the {\it right well}, see definition (\ref{eq.rho_sp}) with $\ell$ replaced by $r$.
After a short calculation, we obtain the integrated two-particle density of states in the quasi-continuum,
\begin{equation}
n^{tp}_{th}(E) = \frac{1}{2} \left( \frac{r^2}{4\pi} E +   \frac{r}{2\pi} \sqrt{2E} \right) \, .
\label{eq.tp_N}
\end{equation}

Two remarks on this result are in order:
Firstly, Eq.~(\ref{eq.tp_DoS}) predicts a second-order polynomial increase of $\rho^{tp}_{th}(E)$ with $r$.
Recalling the behavior of PR in the quasi-continuum limit of the single-particle case [see around Eq.(\ref{eq.gamma_PR})], we deduce that the PR in the two-particle problem should grow with the same functional dependence on $r$. This is confirmed by the fit with a second-order polynomial in Fig.~\ref{fig.PR_i}.
Thus, values of $r\ge1000$ should be well in the quasi-continuum, i.e., suitable to study particle loss.
Secondly, we note in passing that in the limit of an infinitely large well $r\rightarrow\infty$, one finds $n^{tp}_{th}(E)\propto E$.
Not surprisingly, the resulting {\it constant} DOS for two (non-interacting) particles in a one-dimensional system is exactly the DOS of {\it one} particle in two dimensions.

%%=== APPENDIX - WEAK COUPLING ===
\section{Weak coupling regime and survival probability}
\label{app.P_s_w_c}

We here discuss the link between the probability of finding all particles in the left well and the so-called survival probability
\begin{eqnarray}
P_{\rm surv}(t) &=& |\langle\psi(0)|\psi(t)\rangle|^2 \,,
\end{eqnarray}
which denotes the probability of finding the system at a given time $t$ in the initial state $| \psi(0) \rangle$.
The former probability corresponds to $P_1(t)$ in the two-particle case and to $P_\ell(t)$ in the single-particle case.

We begin with the single-particle case for which, according to Eq.(\ref{eq.c_n}), the initial state 
(i.e. an eigenstate $|\chi_i^{\ell} \rangle$ of the isolated left well) can be expanded as 
\begin{eqnarray}
|\chi_i^{\ell} \rangle = | \psi(0) \rangle= \sum_n c_n |E_n^{(sp)}\rangle \, ,
\label{eq.app_c_n}
\end{eqnarray}
where $\{|E_n^{(sp)}\rangle\}$ are the eigenstates of the total system, i.e., of the single-particle Hamiltonian $H_{sp}$ (\ref{eq.H_sp}). 

Next, we remark that our setup fulfills the so-called weak coupling condition between the left well (system) and the right well (environment):  
That is, for the low energies considered here, the projection of every eigenstate $|E_n^{(sp)}\rangle$ onto
the left well is either almost zero or approximately proportional to an eigenstate $|\chi_j^{\ell} \rangle$ of the isolated left well.
The latter implies that the level spacing $\Delta_n=\epsilon^{(sp)}_{n+1}-\epsilon^{(sp)}_{n}$ of consecutive eigenstates of the isolated left well is much larger than the width of the corresponding Lorentzian distribution of the coefficients $|c_n|^2$. 
Otherwise, the Lorentzian distributions of adjacent low-lying levels would significantly overlap. In that case, the projection of an eigenstate $|E_n^{(sp)}\rangle$ (with energy in the latter overlapping interval) onto the left well could yield superpositions of $|\chi_j^\ell\rangle$.

We formulate the weak coupling condition for those eigenstates $|E_n^{(sp)}\rangle$ for which $c_n \neq 0$ as
\begin{align}
E_n^{(sp)}(x)  \approx b_n |\chi_i^{\ell}(x) \rangle \  \text{, for}  \ x \in [0,\ell] \, ,
\label{eq.E_n_prop_c_n}
\end{align}
$b_n$ being the proportionality factor.
From \ref{eq.E_n_prop_c_n} and \ref{eq.app_c_n} follows that $b_n$
approximately equals the expansion coefficient $c_n$
\begin{eqnarray}
c_n \approx b_n \, .
\end{eqnarray}
We highlight two major implications of the latter relation:
On the one hand, we find for each eigenstate $|E_n^{(sp)}\rangle$ of the total system with $c_n \neq 0$:
\begin{eqnarray}
P_{\ell}(|E_n^{(sp)}\rangle) &=&  \int^{\ell}_0 {\rm d}x \  |E_n^{(sp)}(x)|^2 = |b_n|^2 \approx |c_n|^2 \, .
\label{eq.app_D_P_l=c_n}
\end{eqnarray}
If (as in our case) the coefficients $|c_n|^2$ are Lorentzian distributed, so is the probability  $P_{\ell}(|E_n^{(sp)}\rangle)$ (for each non-vanishing component), see Footnote \ref{foot.weak_coupl}.

On the other hand, we can show that $P_{\ell}(t)$ and $P_{\rm surv}(t)$ approximately coincide:
\begin{eqnarray}
P_{\ell}(t) &=&  \int^{\ell}_0 {\rm d}x \  |\psi(t)|^2  \nonumber \\
							 &=&  \sum_{n,m} c_n^{*}c_m \int^{\ell}_0 {\rm d}x \  E^{(sp)*}_n(x)E_m^{(sp)}(x) e^{i(E_n^{(sp)}-E_m^{(sp)})t} \nonumber \\
 							 &\stackrel{(\ref{eq.E_n_prop_c_n})}{\approx}&  \sum_{n,m} c_n^{*}c_m b^{*}_n b_m e^{i(E_n^{(sp)}-E_m^{(sp)})t} \int^{\ell}_0 {\rm d}x \  \chi_i^{\ell}(x) \chi_i^{\ell}(x) \nonumber	\\
							 &\stackrel{(\ref{eq.app_D_P_l=c_n})}{\approx}&  \sum_{n,m} |c_n|^{2} |c_m|^2 e^{i(E_n^{(sp)}-E_m^{(sp)})t}\nonumber\\
							 &=& \big|\sum_n |c_n|^2 e^{-iE_n^{(sp)} t}\big|^2 
							 						 	= P_{\rm surv}(t) \, ,
\label{eq.app_P_l=P_s}
\end{eqnarray}
where we used that the $\{c_n\}$ and $\{b_n\}$ are real numbers, due to the real-valued $H_{sp}$ in configuration space (for finite configuration space basis a real symmetric Hamiltonian matrix).
The last line of (\ref{eq.app_P_l=P_s}) directly implies that (see,
e.g. \cite{HCGK06})
\begin{eqnarray}
P_{\rm surv}(t) = |{\cal FT}[ |c_n|^2 \delta(E^{(sp)}-E_n^{(sp)}) ] |^2 \, ,
\end{eqnarray}
i.e., the survival probability is the (absolute value squared of the) Fourier transform (${\cal FT}$) of the so called local density of states $\rho_{loc}(E) = |c_n|^2 \delta(E^{(sp)}-E_n^{(sp)}) $.
A Lorentzian distribution of the coefficients $|c_n|^2$ (see inset of Fig.~\ref{fig.sp_P_Left_PR}) thus implies 
an exponential decay of $P_{\rm surv}(t)$ and ---in the weak coupling approximation--- an exponential decay of $P_{\ell}(t)$.

For two bosons, the identical calculation is performed by replacing  $\epsilon_n^{(sp)}\rightarrow\epsilon_n$, $E_n^{(sp)}\rightarrow E_n$, $\chi_n^\ell(x)\rightarrow\psi_{k_{1,1}k_{2,1}}(x)$, and $P_{\ell}(t)\rightarrow P_1(t)$.
That is, the relations (\ref{eq.app_D_P_l=c_n}) and  (\ref{eq.app_P_l=P_s}) directly carry over to the two-particle case, for which the survival probability is (almost) identical to the probability $P_1(t)$ of finding both bosons in the left well.

%%%%%%%%%%%%%%%%%%%%%%%%%%%%%%%%%%%%%%%%%%%%%%%%%%%%%%%%%%%%%%%%%%%%%%%%%%%%%%%%%%%%%%
\end{document}